\documentclass[12pt]{article}

\usepackage{pdproc,epsfig} 

\begin{document}

\renewcommand{\thefootnote}{\alph{footnote}}
  
\title{HANOHANO: A DEEP OCEAN ANTI-NEUTRINO DETECTOR\\
   FOR UNIQUE NEUTRINO PHYSICS AND GEOPHYSICS STUDIES}

\author{JOHN G. LEARNED, STEPHEN T. DYE AND SANDIP PAKVASA}

\address{ Department of Physics and Astronomy, University of Hawaii,
 \\ 2505 Correa Road, Honolulu, Hawaii 96822, USA\\
 {\rm E-mail: jlearned@hawaii.edu} }

\centerline{\footnotesize}

\abstract{The science potential of a 10 kiloton deep-ocean liquid scintillation detector for 
~1 MeV energy scale electron anti-neutrinos has been studied.  Such an instrument, designed 
to be portable and function in the deep ocean (3-5 km) can make unique measurements of the 
anti-neutrinos from radioactive decays in the Earth''s mantle. Ths information speaks to some 
of the most fundamental questions in geology about the origin of the Earth, plate tectonics, 
the geomagnetic field and even somewhat indirectly to global warming.  Measurements in 
multiple locations will strengthen the potential insights. On the particle physics side, we 
have identified a unique role in the study of anti-neutrinos from a nuclear power complex, at 
a range of 55-60 km off shore.  Not only can precision measurements be made of most neutrino 
mixing parameters, including $\theta_{13}$ (depending on magnitude), but the neutrino mass 
hierarchy can be determined in a method not heretofore discussed, and one which does not rely 
upon matter effects. This detector is under active study on paper, in the laboratory, and at 
sea. An interdisciplinary and international collaboration is in formation, and plans are in 
motion for a major proposal, to be followed by construction over several years.}

\normalsize\baselineskip=15pt

\section{Introduction and Rationale}

Neutrino studies have made phenomenal progress in the last decade, since the construction of 
kiloton scale specialized detectors, and the good fortune to be able to learn of two neutrino mass 
differences and three neutrino mixing angles.  Of particular interest is the development of large 
detectors for observing the interactions of electron anti-neutrinos ($\bar{\nu_e}$).  Reines and 
Cowan \cite{reines} employed the same technique fifty years ago in making the first neutrino 
observations near a reactor, employing the ``inverse beta'' process whereby a $\bar{\nu_e}$ with 
energy greater than 1.8 MeV is able to change a (free, not nuclear-bound) proton into a neutron 
plus a positron. The positron annihilates immediately providing a flash of light proportional to 
the incoming neutrino energy (less 0.8 MeV), and the neutron wanders about for 100-200 microseconds 
($\mu s$) until it is captured by another proton to make deuterium.  The 2.2 MeV binding energy of 
deuterium gets released in gamma radiation, creating a second flash in the scintillator, of well 
known intensity.  The combination of two pulses, nearby in time and space, plus the constraints 
upon energy, provide a marvelous tool to tag $\bar{\nu_e}$ interactions, discriminating them from 
the huge number of single flashes of similar intensity due to local background and solar 
neutrinos.

Over the past decades, several generations of neutrino detectors have measured neutrinos with 
ever increasing distance from reactors, mostly searching for evidence for neutrino 
oscillations\cite{reactorexpts}.  The electrically neutral neutrino has long been known to be 
very light, perhaps having no mass at all according to many theoreticians.  The peculiarity 
of quantum mechanics allows for the so called ``flavor states'' of neutrinos (electron, muon 
and tauon) to be different than the mass (eigen)states (creatively numbered 1,2 and 3).  In 
this case the waves of different mass states move at slightly different velocities, their 
relative phases change (beat) and the resulting flavor state will thus oscillate with 
distance (for example going from electron to muon to tauon to electron...). In 2002 the newly 
built KamLAND 1000 ton liquid scintillation detector, constructed in a mine tunnel in Japan, 
reported the observation of $\bar{\nu_e}$s from nuclear power reactors spread around Japan at 
a typical range of 180 km\cite{kamland1}.  Most particularly they reported a deficit of 
neutrinos from the expected $\bar{\nu_e}$ flux, which flux had been measured at (no 
oscillation) expected levels repeatedly in previous experiments closer to reactors.  Further 
data collection permitted the observation of not only the disappearance of the $\bar{\nu_e}$, 
but also resolution of the pattern of disappearance and reappearance with energy of the 
neutrino (with periodicity in distance/energy) which characterizes the oscillations 
phenomenon\cite{kamland2}, leaving no doubt about the cause of the neutrino disappearance 
reported earlier. Simultaneously this set of observations coincides with the interpretation 
of the solar neutrino deficits (in electron neutrinos, not anti-neutrinos) observed over many 
years, as uniquely due to oscillations, as indicated by the SNO experiment.

The neutrino energy span from reactors, of roughly 2 to 7 MeV, is due to nuclear fission 
products, some of which are far from the valley of nuclear stability and which can have 
relatively large decay energies. Beta decay neutrinos from natural radioactivity, 
particularly in the decay chains from uranium and from thorium, reach up to about 3.3 MeV. 
People have speculated on the detection of the radioactivity of the whole earth for some 
time, beginning in the 1960's \cite{geonu}.  Even though the flux is rather large (few 
million per $cm^2/sec$ from the whole earth), the cross section for neutrino interaction is 
terribly small (order of $10^{-42} cm^2$), so the expected rate (at Kamioka) amounts to only 
around 26 per kiloton per year in the energy range between 1.7 and 3.4 MeV for these 
geo-neutrinos ("geonus"). The KamLAND group extracted a signal for these geonus and published 
their results in Nature in 2005\cite{kamland3}, marking the first (marginal) detection of the 
earth''s total radioactivity, and opening the door to a new way to study the inaccessible 
deep earth.

\subsection{Genesis of Hanohano Idea}

Several currents of interests steered us towards the idea of building a large version of the 
KamLAND in the deep ocean.  First, the geonus observable at locations such as Kamioka, and 
indeed any location on or near a continental plate arise dominantly from the radioactive 
decays in the proximate crustal rock.  Geochemists believe that the fractional U and Th 
composition of the crust exceeds the mantle by about a factor of 100.  Hence, despite the 
vastly larger mass of the mantle (by also about a factor of 100), the mean distance to 
sources leaves the crustal neutrinos expected to be dominant at a location such as Kamioka by 
70\%/30\%.  For reasons to be discussed further below, the distribution of U/Th in the mantle 
and core capture the most geological interest since they are expected to be the prime source 
of internal earth heating, driving plate tectonics.  As it happens, the oceanic crust is 
thinner and less radioactive, and hence a measurement from a mid-ocean location of mantle 
(and core) neutrinos becomes attractive.  Moreover, the nuclear power reactors contribute one 
of the most serious background sources for geonu measurement, particularly in heavily nuclear 
powered Japan (and France and Eastern US).  One may consider ocean island based detectors, 
but then making measurements at various geographical locations becomes prohibitive.  Thus we 
started to think about a "portable" detector, conceptually an ocean-going KamLAND style tank, 
which could be lowered into the deep ocean, recovered for service and moved to survey new 
locations.

Another current in this thinking came from defense considerations, where growing concern 
about nuclear weapons proliferation has stimulated defense thinkers to contemplate methods 
for remote nuclear reactor monitoring, and even for detection of clandestine nuclear weapons 
testing. These ideas have been tossed about over the years, but people were daunted by the 
inescapably huge size of detectors required to do the job.  We began by playing the game of 
asking how big and how many such detectors would be needed, and in what locations, in order 
to monitor reactors around the world.  This exercise in seeking upper bounds, revealed that a 
total detector mass on the order of a few $km^3$ equivalent would do the job, particularly if 
deployed in units in the range of 10 to 100 megatons, and most effectively distributed near 
land masses and in lakes around the world (about the same number of detectors as reactors, 
~500)\cite{nacw}.  While the 1000 ton KamLAND instrument is the largest scintillation 
detector built so far (2007), there are serious proposals for megaton instruments using water 
as the detection medium (leaping from the spectacularly successful 50 kiloton 
Super-Kamiokande water Cherenkov detector, HyperK, UNO and MEMPHYS). With the addition of a 
dopant material to capture neutrons efficiently (such as $GdCl_3$), these instruments will 
have capabilities for remote reactor monitoring (but not low enough threshold to study 
geonus).  There is also a full $km^3$ sized detector, ICECUBE, under construction at the 
South Pole, albeit with a much higher energy threshold. And plans are moving forward for a 
gigantic deep ocean neutrino detector in the Mediterranean (NESTOR, ANTARES and NEMO joined 
for the KM3 project).  So, one may say that the type of instrument which will be needed for 
a world remote reactor monitoring network, while not practical today, certainly is coming 
over the technological horizon.

Liquid scintillators produce roughly 30 times the light compared to detectors utilizing the natural 
Cherenkov light. The cost for organic based scintillator exceeds that of purified water by several 
orders of magnitude, and simply cannot be afforded when dreaming of 10-100 megaton detectors.  The 
dominant cost factor in this case devolves to the light sensors.  The far future (decades out) 
instruments will need 21st century light detection devices, for which the basic technology exists 
but has not yet been developed. So, for now we are constrained to utilize the well developed old 
glass globe photomultiplier technology, which costs on the order of \$1/cm$^2$ of photocathode 
area.

Various studies have shown that some interesting precision measurements of neutrino mixing 
parameters might be made off-shore from a nuclear reactor complex (more about this below).  
We thus settled into study of a deep ocean version of KamLAND as a geophysics and particle 
physics instrument. We have converged upon a 10 kiloton scale detector for reasons of 
counting rate for important science goals.  The requirements of a geonueutrino detector 
differ by little from those for an off-shore (and necessarily not so deep) instrument for 
particle physics studies. In both cases the radio-purity requirements are not as stiff as for 
(single flash) solar neutrino observations (for which the KamLAND detector is being upgraded, 
and the 200 ton Borexino detector in Italy is being readied for operation in late 2007).

Hence a scaled up version of KamLAND seems to present an excellent target for a next stage 
``terrestrial neutrino telescope'', one employing proven technology, getting experience in 
building such sealed instruments for the deep ocean and offering world class science goals in 
both geology and physics\cite{nusci,doanow}.

First let us describe the technology of the Hanohano detector, and then we will discuss the 
geology (Section 3) and particle physics missions (Section 4) in more detail, discuss 
ancillary missions, the long range prospects and make a summary.

\section{Hanohano Design Studies}

The following are excerpted from the first engineering design report on Hanohano prepared by 
Makai Ocean Engineering in 2006\cite{ceros}, and while necessarily brief, hopefully this 
conveys a sense of the present stage of design and (positive) conclusions about the 
feasibility of a detector on this scale.  The baseline design is as follows:

%%%%%%%%%%%% FIGURE 1 %%%%%%%%%%%%%%%
\begin{figure}[htbp]
%\vglue -3.0cm
\begin{center}
%\hglue -0.5cm
\includegraphics[width=0.5\textwidth]{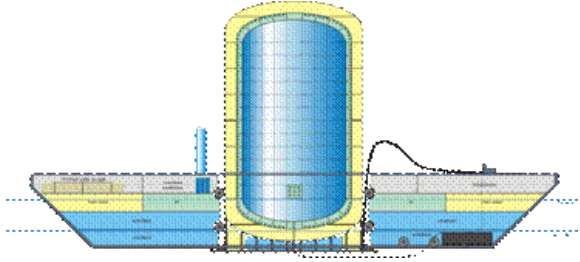}
\end{center}
%\vglue -0.4cm
\caption{A cross sectional sketch of the 10 kiloton Hanohano detector and barge. 
The barge is 112 m long with 32 m beam, and the tank is 26 m in diameter by 45 m tall.}
\label{fig:detector}
\end{figure}
%%%%%%%%%%%% FIGURE 1 %%%%%%%%%%%%%%%

\subsection{Barge}

The support barge has a overall length of 112 m, a beam of 32.3 m (just fitting the Panama Canal), 
and an overall height of 13.8 m. Fully loaded draft is less than 10 m, so that instrumentation can 
proceed at dockside, as illustrated in Figure \ref{fig:detector}. The support barge has a 
conventional barge-shaped bow and stern. With a nominal draft, the detector can be towed safely and 
reliably to any location on the world''s oceans. The barge contains tanks for storing all the 
scintillator and all the oil used in the detector. Multiple stainless tanks are used to maintain 
the liquid purity, to facilitate transfer during (radio-purity) distillation, and to maintain barge 
stability when tanks are partly filled; nitrogen is used in unfilled voids to maintain purity. The 
barge serves as the hotel for detector support. Crew quarters, lab space, generators, water 
production equipment, nitrogen storage, and scintillator cleaning equipment are located in or on 
the barge. The detector uses a nominal $7000~m^3$ of fresh water to eventually flood the veto 
region during submergence. This fresh water is generated with a built-in reverse osmosis plant. 
Producing the fresh water needed is more economical than transporting it. Based upon equivalent 
ship construction costs, the estimated cost of the support barge is 9 million dollars.
 
\subsection{Detector Tanks}

\begin{figure}[htbp]
%\vglue -3.0cm
\begin{center}
%\hglue -0.5cm
\includegraphics[width=0.5\textwidth]{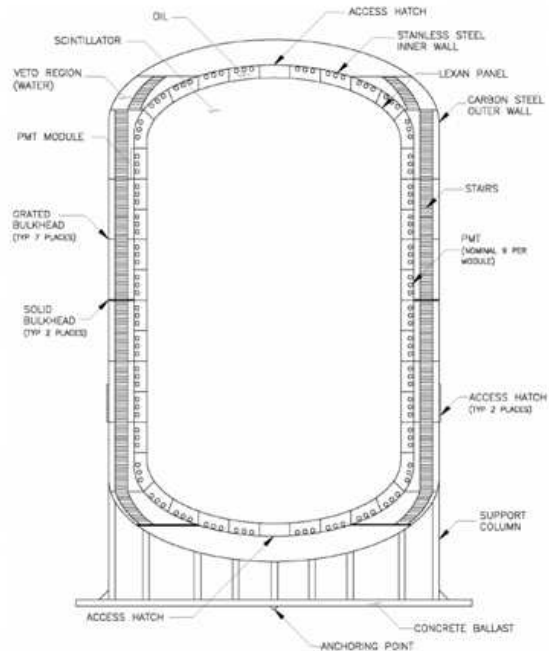}
\end{center}
%\vglue -0.4cm
\caption{A cross sectional sketch of the 10 kiloton Hanohano detector tank which is 26 m in 
diameter by 45 m tall, has scintillator in the inner volume, oil around the light detectors, and 
pure water in the outer veto region.  Optical detectors are in clusters of 9, and all connections
are in the veto region under oil.}
\label{fig:tank}
\end{figure}

The detector is cylindrical with an outer diameter of 26 m, and an overall height of 44.7 m, as 
shown in Figure \ref{fig:tank}. The concentric fiducial volume has a diameter of 20 m and an 
overall height of 35 m. There is a buffer layer 1 m thick surrounding the scintillator and a 2 m 
layer of fresh water veto region surrounding that layer. There are approximately 4300 phototubes 
spaced at a nominal 0.8 m in the buffer region surrounding the fiducial volume. Clear Lexan plates 
separate the scintillator region from the surrounding oil. The 2 m wide veto region is built in 
layers with horizontal grated decks spaced at a nominal 4.8 m. Decks are accessible within the 
detector through stairwells in the veto region. The veto region is the main manned access into the 
detector, with all wiring and plumbing going through this area. Some decks within the veto region 
are water tight, to provide stable and minimal free-surface flooding stages during deployment.

Large polypropylene compensator bags are located at the bottom of the detector to compensate for 
volume changed due to compression and cooling (to ~4$^o$ C) at 4000 m depth. The compensator bags 
compensate for the fresh water and scintillator only. The oil buffer region is compressed by a 
flexing of the Lexan cover plate; therefore, the oil volume change is compensated by larger 
scintillator fluid compensation bags. Compensation is about 5\% of the total volume of the detector. – 
Half of that is due to cooling, half due to compression. Pre-cooling the scintillator and oil was 
considered, but this did not prove to be economically viable.

The detector has a nominal concrete ballast of 744 tonnes (wet). This will vary as the structural 
requirements are refined. The detector has an anchor weight of 607 tonnes (wet). When the detector 
is launched, it weights 86 tonnes (wet), including the anchor. When it drops its anchor and leaves 
the bottom after a lengthy deployment, it is 87 tonnes buoyant. Transit times up and down are 38 and 
39 minutes respectively (as revealed by computer simulation).

%%%%%%%%%%%% FIGURE 1a %%%%%%%%%%%%%%%
%%%%%%%%%%%% FIGURE 1a %%%%%%%%%%%%%%%

Each PMT compartment contains 9 PMTs in a 3x3 array. These compartments are kept isolated from each 
other in order to (1) avoid differential pressures across the Lexan plate due to density variations 
between the oil and the scintillator and (2) to isolate damage to a single compartment in case there 
is a PMT sphere implosion. (If there is an implosion, and the Lexan plate leaks, only the oil from 
one compartment will mix into the scintillator,  a contamination level that is not desirable but 
acceptable to the physics.) All plumbing for flooding and draining the interior of the detector is 
contained within the veto region.  All connections will be made in an oil bath, greatly enhancing 
reliability. The estimated cost of the detector is \$29 million. Deployment would be an 
additional \$2 million.  Additional costs bring the complete detector construction into the \$50 
million range.  Usually project costs will multiply this by a factor of three to five in total cost 
accounting.

\subsection{ Technical Feasibility Assessment}

Hanohano's critical components relative to operation in the deep-ocean, massive construction, and 
deployment have been investigated and there are no show-stoppers. Key critical issues such as 
scintillator performance in the deep-ocean, construction and deployment, low-power operations and 
control, failure survivability, and electronic design have been investigated, prototyped and/or 
tested. At this time, this detector is a practical and technically feasible method of detecting 
anti-neutrinos.

Hanohano size:  Studies of detector counting rates and backgrounds have driven the size goal 
from an initial 1.4 to 4, to 10 k-tonne. The basic requirement was to make useful 
measurements in one year, with the option of relocation on a roughly annual basis. Geologists 
have flagged important reasons for measurements at various locations in the world's oceans. 
The size increase also facilitates the ability to make unique and important measurements of 
neutrino properties. From the science standpoint, of course, larger is always better for more 
statistical power. It seems that the 9-10 kiloton range is adequate for the geological and 
elementary particle science.  Detectors up to about 120 kiloton were found to be practical, 
but our engineering judgment is to stay at the lower (10 k-tonne) size for the first 
instrument.

Structure design:  A variety of anti-neutrino detector shapes and configurations were 
analyzed for deployability, stability, and structural integrity resulting in the elimination 
of most concepts and shapes and focusing on a cylindrical detector that is supported by a 
separable surface barge. This concept was optimal for a very wide range of detector sizes 
from 2 k-tonnes to 100 k-tonnes.

PMT design:  At this point the only viable photodetector for large scintillation detectors is 
the classic glass bulb photomultiplier tube (PMT). Due to the hundreds of atmospheres of deep 
ocean pressure, these PMTs must be protected by a glass instrument housing. There are four 
projects presently utilizing such detectors, and so viable, well studied hardware is 
available for a starting point in the Hanohano design. At present a 10-inch diameter PMT in a 
13-inch housing is favored, which is very similar to the (thousands of) units being installed 
in the deep ice at the South Pole in the ICECUBE detector.

Implosion risk:  We have analyzed and demonstrated a practical method of altering the pressure wave 
and preventing sympathetic implosions in the closely-spaced instrument housings needed in these 
detectors. This greatly reduces the perceived and real risk of operating this detector in the deep 
ocean.

Low Power operation:  We have examined the utilization of modern low power integrated electronics to 
the end of achieving sufficiently low power to enable one year operation on battery power alone. We 
have made an order of magnitude progress over older designs, and can foresee another order of 
magnitude gain in future design iterations. At present, battery operation is marginally possible. 
However, for presently planned operations with fiber optic cabling to shore there will be no problem 
in handling the $<$2kW we estimate for the 10 kiloton detector.

Deployment:  The Hanohano detector and support barge have been designed specifically for deployment 
simplicity and reliability. The detector can be built such that the scintillator and oil buoyancy 
supports the structure and makes deployment economically viable. Cable attachment and lays to shore 
are well within cable laying state-of-the-art technology.

Scintillator test and selection:  We have examined various candidates for scintillation fluid and 
find that a commercially available liquid (LAB, a precursor to dish-soap), will meet our needs. 
Experimental studies of temperature and pressure dependence reveal no stoppers. Further work is 
needed for optical characterization and studies of the level of necessary filtration.

Internal communications:  A plan for a tree structure of internal connection has evolved which 
evades system vulnerability from single point failure. Cables will be used in the first level from 
PMT to digitizer, and fiber optics thereafter, with a redundant path to shore. The design minimizes 
at-sea data processing and takes advantage of the ongoing advances in submarine fiber optic 
communications to send all data to shore for trigger recognition, event building and filtering, 
reconstruction and storage.

Summary of Conceptual Plan:  A conceptual design of the $10,000~m^3$ scintillator Hanohano 
detector and barge has been completed that will just fit through the Panama Canal. Weight, 
stability, structural needs, operational requirements, and physics needs have been checked 
and are reasonable; the conceptual vehicle is a feasible approach.

\subsection{ Further Study}

There remains much work to be done in detailed design.  We highlight some studies needed to 
converge upon final engineering requirements, aside from second iteration in all design 
areas:

Scintillator: We are fortunate that great progress has been made in recent years in the 
development of techniques for removing radioactive materials from liquid scintillators. These 
techniques can be used with Hanohano. The scintillator will be purified by four methods: 
water extraction, nitrogen stripping, vacuum distillation, and filtration. More laboratory 
studies are needed to converge on the optimal scintillator cocktail.

Ocean Demo:  A small demonstration model has been conceptually designed for specific and 
critical next-step testing. This testing will include an ocean deployment of about a one ton 
module, large enough to demonstrate the radio-purity, protection from background, and the 
functionality of the optical modules.

Science Simulations:  In the study of the detector and science applications we have reached 
the happy conclusion that the initial notion for a deep ocean scintillation detector, 
enlarged to the 10 kiloton scale, will be a detector of world importance for science with a 
wide interdisciplinary program cutting across geology and physics.  Further work is required 
in more detailed computer simulations, particularly in the optimization of the size and 
number of optical modules required.  An outstanding issue for the particle physics 
measurements relates to the required energy resolution for achieving the Fourier transforms 
of oscillations as discussed below.

\section{Geology}

\begin{figure}[htbp]
%\vglue -3.0cm
\begin{center}
%\hglue -0.5cm
\includegraphics[width=0.5\textwidth]{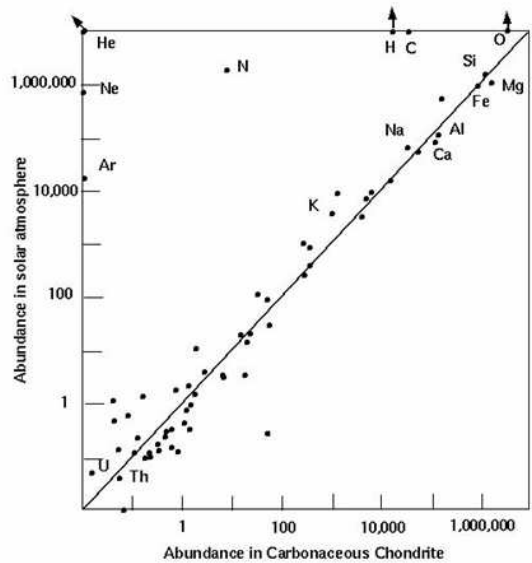}
\end{center}
%\vglue -0.4cm
\caption{Comparison of the relative abundances of atomic species
in the solar atmosphere versus that found in a Carbonaceous 
Chondrite \protect\cite{mcdonough}. }
\label{fig:abundances}
\end{figure}

Key questions in geology are where are the uranium and thorium in the Earth and what do we 
really know about the chemical composition of the inner earth? The answer to the latter is 
unfortunately, not very much, and in fact perhaps less than we know about the inside of the 
sun. We can only guess the overall composition of Earth by analogy, using spectroscopy of the 
outer sun and direct measurements of meteorites\cite{mcdonough}. In fact only three 
Carbonaceous Chondrites are generally taken to provide the template for terrestrial 
composition, see Figure \ref{fig:abundances}.  Of course, we can directly observe only the 
materials at or near the Earth''s surface. There are expected and modeled differences in this 
composition and the proto-earth abundances, due to early heating which drove off light 
elements, and due to chemical combinations of some elements which may be shallower or deeper 
within the earth. From our physicists view it is a complicated story, without even a 
consensus upon the earth formation sequence, which certainly presents multiple possible 
scenarios.

%%%%%%%%%%%% FIGURE 2 %%%%%%%%%%%%%%%
%%%%%%%%%%%% FIGURE 2 %%%%%%%%%%%%%%%

\begin{figure}[htbp]
%\vglue -3.0cm
\begin{center}
%\hglue -0.5cm
\includegraphics[width=0.5\textwidth]{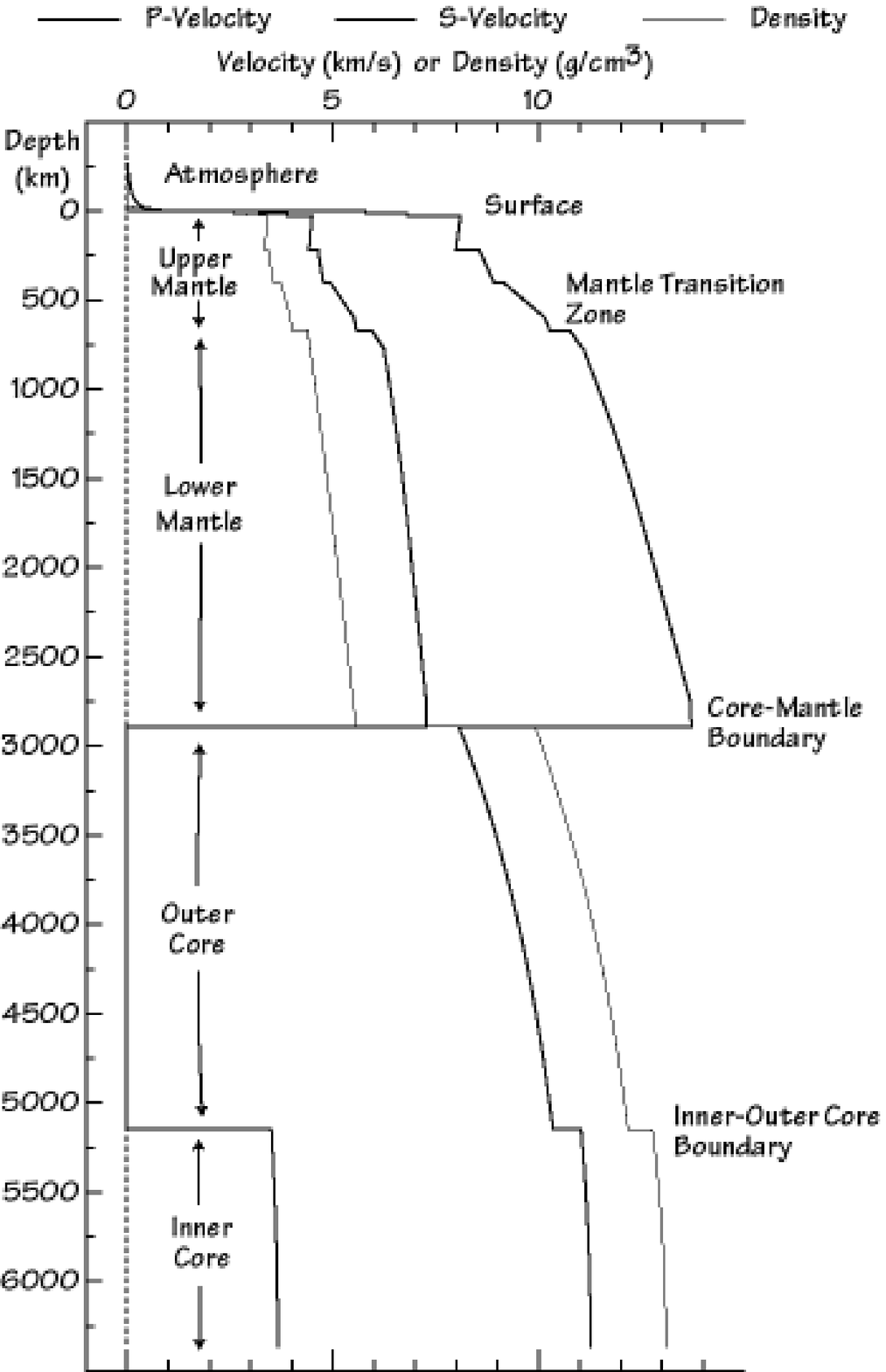}
\end{center}
%\vglue -0.4cm
\caption{The velocity and inferred density profiles of the earth as presented in 
the Preliminary Reference Earth Model of Dziewonski and Anderson\protect\cite{prem}.}
\label{fig:prem}
\end{figure}

The most direct evidence for the structure of the Earth comes from seismic measurements. 
Multiple recordings of earthquakes yield sound velocity profiles of the earth and even some 
detail on lateral heterogeneity. Combining these with measured earth mass moments (from 
satellites), and an equation of state, one may infer the Earth''s radial densiy 
profile\cite{prem}, as illustrated in Figure \ref{fig:prem}.  However, the composition cannot 
be inferred uniquely from this; one can only posit a possible mixture which would satisfy the 
velocity constraints.

%%%%%%%%%%%% FIGURE 3 %%%%%%%%%%%%%%%
%%%%%%%%%%%% FIGURE 3 %%%%%%%%%%%%%%%

The internal terrestrial heat gradient drives slow mantle circulation, producing continental drift, 
seafloor spreading and tectonic activity. Also the geomagnetic field is thought to originate from the 
circulation of the liquid outer iron-nickel core.  From the frozen-in magnetic fields of dated rocks on the 
surface, we know that geo-magnetic fields have been around for billions of years, and though fluctuating in 
direction have had reasonably constant magnitude. So heat flow from the inner earth has been reasonably 
constant on a billion year timescale.

Earth surface heat outflow is tiny compared to solar irradiance and measurements are difficult, 
particularly from the ocean floor. Data with model-dependent interpolation gives ranges from 30-45 
TW for the total heat emission.  Given expectations on the U/Th content, the U/Th radiogenic heat 
may be in the range of 20 TW, but could be twice that\cite{earth_heat}. Many other possible sources 
of heat have been suggested, but radioactivity is thought to be dominant, though the heat budget 
remains uncertain to a factor of two, see Table \ref{tab:heat}.

The big question is not simply how much U/Th resides in the earth, but whether it has mostly floated 
like slag up under the crust (as most experts believe), or remains in solution in the mantle, or has 
sunk onto the core-mantle boundary, or in a minority opinion even sunk into the core (and 
combinations of all of the above).  One controversial model by Herndon \cite{herndon}, has 
enough U in the inner core to power a natural breeder reactor providing 1-10 TW from the inner core. 
(This geo-reactor, if it exists, will be easy to detect in the new experiments discussed below.)  
While most geologists do not accept this geo-reactor model, it is not at all certain where the U/Th 
resides in the earth. Where the U/Th delivers the radiogenic heating makes a large difference, even 
without a geo-reactor, since presumably the circulation of liquid outer core is the region of origin 
of the geomagnetic field. It would seem that one would need a fire under the pot (the liquid outer 
core) to drive the presumed geo-dynamo. And, one would imagine that the mantle would do well with 
the heat from below, though there are some who argue that the circulation can originate in dropping 
cooling flows.

Another issue has to do with the content of potassium, in particular the radioactive isotope potassium-40 
(K40).  The earth seems to be somewhat depleted in K, relative to external reference abundances, and models 
have been made suggesting that it may have disappeared due to volatility.  However, the inner earth core 
does seem to have slightly less density (based upon seismic velocity) than from the expected nickel-iron 
mix.  Some speculate that the decreased density is due to K in solution (K-S?), which might then allow for 
the K heating to be the real pot-boiler for the outer core from below. Unfortunately the K40 neutrinos are 
of very low energy, and do not make the signature inverse beta decay reaction (not enough energy to promote 
a proton to a neutron plus positron).  Particle physicists have not found any viable plans as yet to 
measure the K40 neutrinos.

\begin{table}[h]
\caption{Summary of terrestrial heat sources, total\protect\cite{earth_heat}.}
\label{tab:heat}
\small
\vspace*{13pt}
\begin{center}
\begin{tabular}{||l|c|c|c||}
\hline \hline
Element/Source     & Abundance(ppm)   & Calc. Heat (TW) & Meas. Heat (TW)   \\
\hline
 Potassium (K)     &  170             &  3.7 $\pm$ 50\%   &         \\
\hline
 Uranium (U)       &  0.018           & 10.0 $\pm$ 50\%  &         \\
\hline
 Thorium (Th)      &  0.065           & 10.5 $\pm$ 50\%? &         \\
\hline \hline
 Total Radioactive &                  & 24.2 $\pm$ 50\%?  &         \\
\hline
 Other Sources     &                  &  $<$10 ?      &         \\
\hline
 Geo-Reactor       &                  &   0-10 ?      &  $<$20  \\
\hline \hline
 Total Heatflow    &                  &  30-50        &   30-45 \\
\hline \hline
\end{tabular}
\end{center}
\end{table}

\subsection{ Natural Neutrino Spectra}

As illustrated below in Figure \ref{fig:KL_spectra}, the dominant fraction of the reactor signal as 
observed by KamLAND, is in an energy region between about 2.0 and 7.0 MeV neutrino energy, 
corresponding to 1.2 to– 5.2 MeV in the observed first pulse energy in the detector. The decay 
energies attributable to uranium-238 decay chain and to thorium-232 decay chain are all below 3.6 
MeV.  There is an additional background show in Figure \ref{fig:KL_spectra}, due to a contamination 
of the KamLAND detector by radon and a reaction of alpha particles with 13C.  This is an avoidable 
background and will not be a factor in later measurements, though it was a nuisance in the initial 
KamLAND attempt at measuring the U/Th neutrinos, as reported in the cover issue of Nature in July 
2005\cite{kamland3}.

\begin{figure}[htbp]
%\vglue -0.50cm
\begin{center}
%\hglue  -0.5cm
\includegraphics[width=0.5\textwidth]{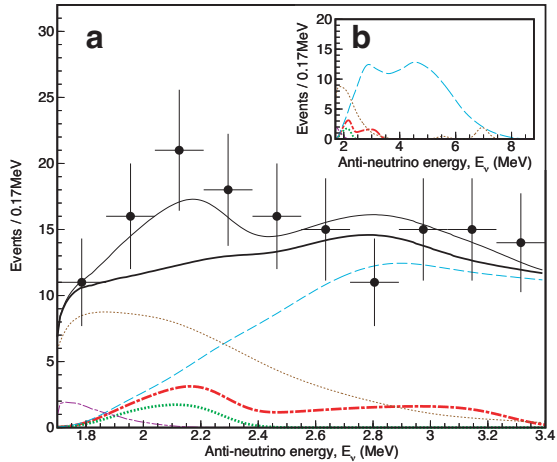}
\end{center}
%\vglue -1.8cm
\caption{Neutrino spectra, predictions and data at KamLAND, showing both geoneutrinos and reactor 
neutrinos, plus backgrounds. The reactor flux is indicated by the dashed curve in panel a) and b). The 
uranium chain neutrinos are shown by the heavy dot-dash line in panel a), and the thorium by the short
dashed line just below.  The dotted curve accounts for a contamination background due to Radon. The the 
heavy solid line represents the summed sources other than geo-neutrinos, and the light solid line shows 
the total of all contributions and along with data points\protect\cite{kamland3} .}
\label{fig:KL_spectra}
\end{figure}

\begin{figure}[htbp]
%\vglue -0.50cm
\begin{center}
%\hglue  -0.5cm
\includegraphics[width=0.5\textwidth]{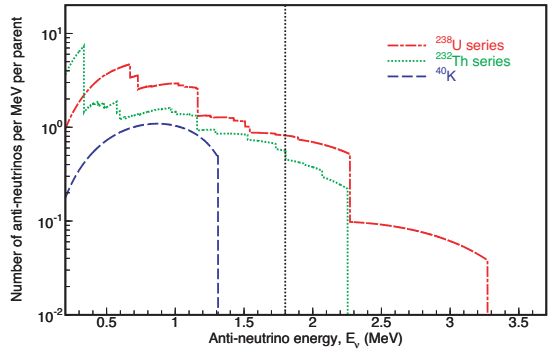}
\end{center}
%\vglue -1.8cm
\caption{Geoneutrino spectra from the uranium (dot-dash line) and thorium (dotted line)
decay chains, plus that from potassium-40 (dashed line)\protect\cite{kamland3}.}
\label{fig:geonu_spectra}
\end{figure}

One may also note that the spectra from U and Th differ significantly, so with adequate statistics 
we can measure the ratio of U/Th as well as observe the total flux and hence amount of U and Th, as 
illustrated in Figure \ref{fig:geonu_spectra}. Observing the total rate does not correspond exactly 
to the total abundance of U and Th however, even in a uniformly layered earth (it is not the same 
as for electrical charges and Gauss' 'Law).  Moreover there are surely great lateral 
heterogeneities due to the varying crustal composition.  Most U/Th is expected to be in or near the 
crust, so discerning the amount distributed throughout the mantle and core is very difficult.  For 
example, only 25\% of the flux from U/Th decay at KamLAND is expected to be from the mantle and 
core, and most due to the local mountains and deeper plate.  Oceanic crust is younger and thinner 
and expected to have typically only 1/10 as much U/Th as that when measuring from a continental 
location, and hence the crucial issue of how much of the terrestrial radioactivity is in the 
mantle/core will need to be measured from an oceanic location.

%***********************************************************************************************
%***********************************************************************************************

\subsection{ Can We Measure More Than Rate?}

Of course we would like to measure the arrival directions for the neutrinos and hence map out the 
origin in direct fashion.  However, directional measurement is very hard at these energies and 
particularly in a scintillating material (which gives off light isotropically).  A small handle can 
be had from the net momentum transmitted to the neutron by the neutrino, which statistically biases 
the locations of the positron annihilation and the neutron capture to be slightly aligned with the 
original neutrino direction.  The Chooz reactor experiment in France did achieve a 18 degree 
resolution from their nearby reactor with several thousand events\cite{chooz}.

\subsection{ Synergy in Measurements at Various Locations}

It is important to have more than ocean measurements of the anti-neutrino flux.  We do not know well the 
total U/Th content of the earth, nor do we know well the location of the material.  Since the land-based 
measurements are dominated by the relatively local crustal flux, they still must subtract the mantle (and 
core, if any) contributions which are not negligible (10-30\%).  For the ocean measurements as well, the 
crustal components make contributions to be subtracted to extract the pure mantle/core flux. However, since 
measurements on the continents will vary substantially (mostly due to crust thickness), it is vital to have 
multiple locations.  Without belaboring the point, there is synergy in multiple locations as we set about 
untangling the location and distribution of the U/Th throughout the earth.

\begin{table}[h]
\caption{ Some proposed geo-nu instruments, location, size and status.}
\label{tab:expts}
\small
\vspace*{13pts}
\begin{tabular}{||l|c|c|c|c||}\hline \hline
Project    &  KamLAND    &  Borexino   &   SNO+      &   Hanohano  \\
\hline
Location   &   Japan     &    Italy    &  Canada     &   Hawaii    \\
Crust      & Continental & Continental & Continental & Oceanic \\
Current status or Start date & Operating & 2007 & 2008 & Planning \\
Depth (meters water equivalent) & 2700 & 3700& 6000 & 4500 \\
Target ($10^{32}$ free protons)& 0.35 & 0.18 & 0.57 & 8.7 \\
Geo-neutrinos per year- Total &13 & 8 & 30 & 110 \\
Geo-neutrinos per year- Mantle & 4 & 2 & 5 & 81 \\
Reactor neutrinos per year & 39 & 6 & 32 & 12 \\
\hline \hline
\end{tabular}
\end{table}

The Table \ref{tab:expts} shows some parameters for several operating and proposed geonu detectors.  
Not included are two European based and one Russian proposed instruments.  The LENA 
detector\cite{lena} has been discussed for several locations (in Finland, France or possibly the 
Mediterranean), and with several possible sizes (50-100 kiloton), as a large horizontal 
scintillator tank.  A completely different approach is being taken by a Dutch led group, in a 
program called EARTH\cite{earth}, which would locate directional detectors in drilled holes beneath 
the island of Curacao in the Caribbean. There has also been a large liquid scintillation detector 
suggested for the Baksan neutrino detector facility in the Caucasus Mountains in Russia. We do not 
have enough definite information to include these in the table above.

In sum, Hanohano can make the first definitive measurements of the U/Th content of the earth's 
mantle within one year.  Moving the detector to other locations can then begin to understand the 
possible lateral variation in the amount of U/Th, as may come about from upwelling, or subduction. 
The ratio of U/Th may change significantly too, since the solubility of U and Th is different, and 
crust is being subducted and emerging from mid-ocean ridges.  As yet there are not many detailed 
models since the constraining information has been so indirect.  Geologists have told us that 
measurements even within a factor of a few will be useful; we can envisage getting totals and 
ratios to the 10\% regime, with sensitivity to lateral variation in that scale as well, within a 
few years.  This will open a new line of inquiry into the fundamentals of geology.

\section{Particle Physics}

Over the past eight years we have witnessed the astonishingly rapid realization of neutrino 
oscillations taking place. This began with the Super-Kamiokande observations of muon neutrino 
oscillations, followed quickly by the Super-Kamiokande and SNO observations of solar electron 
neutrinos and the KamLAND observations of electron anti-neutrinos from reactors. These beautiful 
results have convinced the community of the reality and surprisingly large magnitude of neutrino 
mixing.  We have moved from an era of not knowing about non-zero neutrino mass into one of 
excitedly attempting to measure the complete three neutrino mixing scheme, the co-called MNSP 
matrix.  Attempts continue to seek evidence for more than three neutrino types have so-far proved 
fruitless. (The April 2007 preliminary report of the Mini-BOONE experiment refuted the 
controversial results of the LSND experiment, at least as interpreted as due to oscillations 
involving a sterile neutrino).  Hence in order to better understand the peculiar nature of the 
neutrino mixing we must fill in the MNSP matrix, a task surprisingly well underway.  A detailed 
description of the state of the art in 2004 can be found in the U.S. Academy of Sciences White 
Paper \cite{nu_white_paper} and recent reviews, eg.\cite{mohapatra}.

The MNSP matrix can be described for three neutrino oscillation purposes by one phase and three 
mixing angles.  From the experiments reported as of this time we have rather good knowledge of the 
mass differences (though not the order, or ``hierarchy'' as it is called), and we have a respectable 
knowledge of the mixing angles.

So far, with not terribly precise data sets (5-10\% is a typical error magnitude), it has been 
convenient to describe the observed neutrino mixing in terms of effective two-neutrino states.  
This case obtains for both the atmospheric muon neutrino mixing (with tau neutrinos, on a scale of 
GeV in energy and distances of thousands of kilometers) and for solar and reactor neutrinos (the 
latter on a scale of a hundred kilometers and a few MeV).  As it turns out the mass-squared 
differences measured are $\delta {m^2}_{solar} \simeq (7.9 \pm 0.7) eV^2$ and $\delta {m^2}_{atmos} 
\simeq (2.5 \pm 0.5) \times 10^{-3} eV^2$. They are just about a factor of about thirty apart, or a 
factor of 5-6 in mass. (Of course we do not yet know the offset, if any, of the smallest of these 
masses from zero, but which is not more than 0.5 eV).

%%%%%%%%%%%% FIGURE 1 %%%%%%%%%%%%%%%
\begin{figure}[htbp]
%\vglue -3.0cm
\begin{center}
%\hglue -0.5cm
\includegraphics[width=0.5\textwidth]{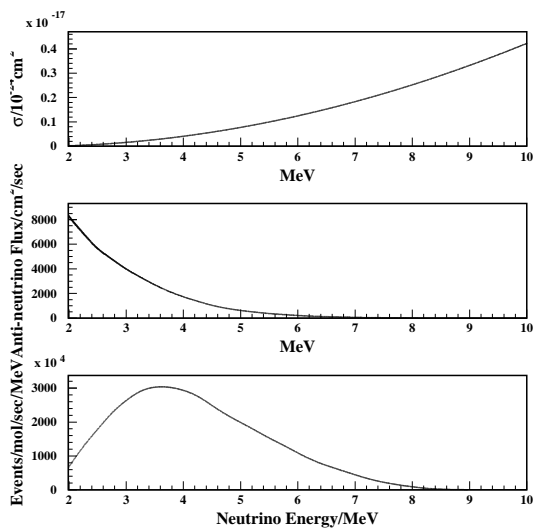}
\end{center}
%\vglue -0.4cm
\caption{Reactor flux, neutrino inverse beta cross section and 
event rate versus anti-neutrino energy. }
\label{flux}
\end{figure}
%%%%%%%%%%%% FIGURE 1 %%%%%%%%%%%%%%%

Mixing angles are generally harder to measure with precision, since they depend upon 
experimental measurements of rates.  This is particularly painful for the case of the 
electron neutrinos as one is restricted to electron neutrino disappearance and depends upon 
otherwise predicted absolute efficiencies, fluxes and cross sections. The muon neutrino case 
is a bit easier because ratios can be employed to dampen systematic influences.  
Surprisingly, the atmospheric neutrino mixing angle ($\theta_{23}$) turns out to be close to 
45$^\circ$.  On the other hand, the solar mixing angle ($\theta_{12}$), seems to be nearer to 
32$^\circ$, and apparently not 45$^\circ$.  The third angle is unknown, though we know it is 
not large.  The limits most generally accepted for the third angle come from the Chooz 
experiment, which found $\theta_{13} < 12^\circ$. This angle could be zero, according to 
present knowledge nothing prevents it from being so. If $\theta_{13}$ is zero or very small, 
it will be impossible to measure CP violation phase in the neutrino MNSP matrix, and learn 
about the possible connection to the matter-antimatter asymmetry in the universe. This has 
made the measurement of $\theta_{13}$ a focus of considerable world effort.  Herein we 
describe a somewhat different approach than discussed (as far as we know) previously, 
employing Fourier transforms in the neutrino data analysis.

First off, we must be careful to use a full three neutrino formulation of the oscillations.  
Most authors make simplifying assumptions ($\delta {m^2}_{23} \simeq \delta {m^2}_{13}$ in 
particular), about which we must now be a bit more precise.  The exact three neutrino formula 
for electron anti-neutrino survival probability can be written as\cite{3nu1,3nu2}:

\begin{eqnarray} 2(1-P(\nu_e \rightarrow \nu_e)) =
 \cos^4 (  \theta_{13}) \sin^2(2 \theta_{12})(1-cos(\Delta_{12}))+ \nonumber \\
 \sin^2 (2 \theta_{13}) \cos^2(  \theta_{12})(1-cos(\Delta_{13}))+ \nonumber \\
 \sin^2 (2 \theta_{13}) \sin^2(  \theta_{12})(1-cos(\Delta_{23})). 
\label{eqn:Pee}
\end{eqnarray}

where $\Delta_{ij} = 1.27 \frac{|{m^2}_i - {m^2}_j |L}{2 E_{\nu}}$, with $m^2$ in units of 
$eV^2$, L the neutrino flight distance in meters, and $E_\nu$ is the neutrino energy in MeV, 
and $\Delta_{23} = \Delta_{13} \pm \Delta_{12}$. This expression applies for the ``normal'' 
hierarchy of masses ($m_3 > m_2 > m_1$) with a minus sign between $\Delta_{13}$ and 
$\Delta_{12}$ or a plus sign for the ``inverted hierarchy''.  In any event the first term is 
responsible for the solar deficit and the modulation of the reactor spectrum observed with 
KamLAND.  It has a wavelength of about 120 km at the peak of a reactor spectrum around 3.5 
MeV (or equivalently about 2.5 kHz).

We write it this way to exhibit clearly the ``periodic behavior'' of the survival probability 
as a function of neutrino range and energy.  The first term dominates, and is largely what is 
measured by the reactor and solar experiments so far (where we assume equality of the 
parameters for electron neutrinos and anti-neutrinos, which at this time is consistent with 
results).

\begin{figure}[htbp]
%\vglue -0.50cm
\begin{center}
%\hglue  -0.5cm
\includegraphics[width=0.5\textwidth]{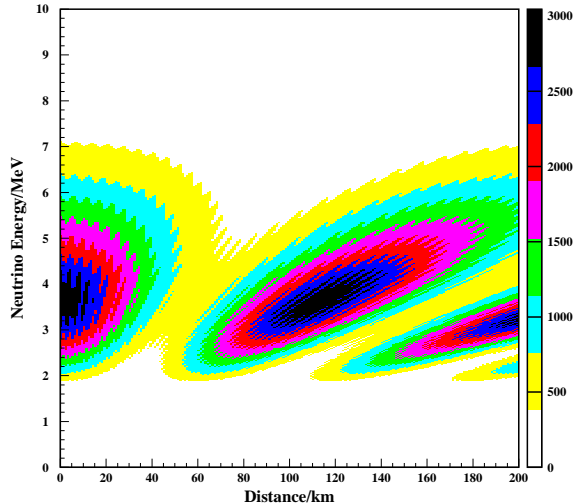}
\end{center}
%\vglue -2.0cm
\caption{Rate versus energy and distance, with $\sin^2 (2\theta_{13}) = 0.1$. Note long distance 
modulation at the solar wavelength, and short scale modulation at the atmospheric wavelength due to 
finite $\theta_{13}$. Note also the lines of constant phase radiating from the origin(at constant 
L/E).}
\label{red}
\end{figure}

We will first discuss the measurement of $\theta_{13}$ and the determination of the neutrino 
mass hierarchy, which are new topics first discussed for application to remote measurement of 
reactor neutrinos for Hanohano.  After this we will return to the more `bread and butter' 
measurement of $\theta_{12}$, and then summarize some of the many other physics measurements 
to be made with Hanohano.

%%%%%%%%%%%%%%%%%%%%%%%%%%%%%%%%%%%%%%%%%%%%%%%%%%%%%%%%%%%%%%%%%%%%%%%%%%%
\section{Measuring $\Theta_{13}$ }
\label{t13}
%%%%%%%%%%%%%%%%%%%%%%%%%%%%%%%%%%%%%%%%%%%%%%%%%%%%%%%%%%%%%%%%%%%%%%%%%%%

One sees that the latter two terms in Equation \ref{eqn:Pee} contain two different 
frequencies, producing a beating phenomenon.  If the solar angle were 45$^\circ$ then the 
beating would be maximal.  The wavelength of this oscillation is only about 4 km at 3.5 MeV, 
and this is what the near-in reactor experiments (Double Chooz\cite{2chooz} in France, Daya 
Bay\cite{dayabay} in China and Kaska\cite{kaska} in Japan) are aiming to detect.  The 
amplitude is however controlled by the $\sin^2 (2 \theta_{13})$ factor which we already know 
to be small, so this is the measurement challenge.  The beating phenomenon is not easy to 
observe because the two factors in the second and third terms in Equation \ref{eqn:Pee} 
differ by about a factor of two to three.  Mostly the higher spatial frequency is dominated 
by the second term and hence by $\Delta_{13}$. Now one may well ask which mass difference is 
being measured by the atmospheric neutrinos experiments.  The problem has not arisen in 
practice as yet because the precision of measurement of the atmospheric mixing is still not 
better than 10\% at the 1 $\sigma$ level and the difference between the two possibilities is 
only about 3\%. In practice, the atmospheric muon neutrino oscillation measured by 
Super-Kamiokande yields some sort of average between the two, a problem we shall not deal 
with in this article, though we will assume for pedagogical purposes that the answer is well 
known.

In an infinitely long data set in $L/E$ (that is to say a huge range in energy at a fixed 
distance, or a combination of spectra from various distances) we would see a Fourier 
transform (in L/E) which had three spikes, at $\Delta_{12}$, $\Delta_{13}$ and at 
$\Delta_{12} - \Delta_{13}$, with amplitudes in the ratio of $13.5~:~2.5~:~1.0$ for 
$\theta_{12} = 32^\circ$ and $\theta_{13} = 13^{\circ}$. The $\Delta_{13}$ peak will be split 
from the $\Delta_{23}$ peak by a mere 3\%, and thus one would imagine to be unobservable in a 
data set less than about 30 periods long in the $\Delta_{13}$ modulation. However, notice 
that with infinite resolution (infinite numbers of events) even over a finite period one may 
overcome this simple restriction (this is an example of the Shannon-Hartley theorem).  
Moreover, one may beat the restriction by employing knowledge of other parameters, such as 
the phases and ``envelope function'' in Fourier terms, or spectrum in our case.  We shall 
discuss this further below.

\begin{figure}[htbp]
%\vglue -0.50cm
\begin{center}
%\hglue  -0.5cm
\includegraphics[width=0.5\textwidth]{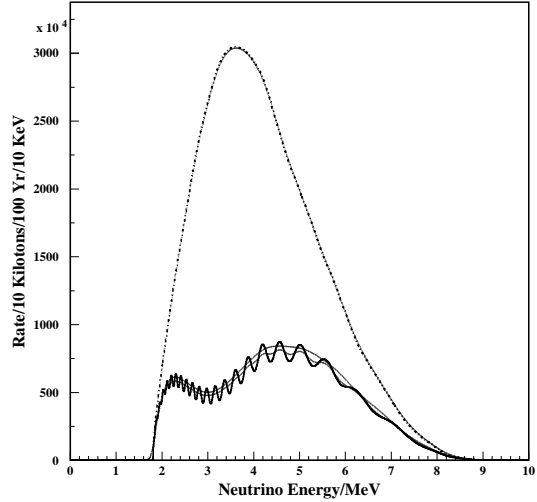}
\end{center}
%\vglue -1.8cm
\caption{Rate versus neutrino energy at a distance of 50 km, 
 with no oscillations, oscillations with $\theta_{13} = 0$ and 
 with $\sin^2(2\theta_{13}) = 0.1$.}
\label{spectrum}
\end{figure}

In Figure \ref{flux} we show the cross section, reactor flux (in this example the flux from 
the San Onofre facility at full power and with a fresh fuel load) and event rate per mole of 
hydrogen.  In Figure \ref{red} we illustrate the spectrum of events versus distance.  Here 
one sees the large hills due to the $\Delta_{12}$ modulation with the smaller and narrower 
$\Delta_{13}$ oscillations superposed. Note the rills in the plot, which point back to the 
origin ($L = 0$ and $E_\nu = 0$), which are lines of constant $L/E$.

One should observe that given a finite reach in energy, namely from about 2-7 MeV from a 
reactor, one will intercept an increasing number of cycles in the $\Delta_{13}$ oscillations 
as one goes further from the reactor.  This might encourage one to take such an experiment to 
great distance in order to observe many cycles and sharpen the Fourier transform.  Of course 
one gets beaten by lower rates, and eventually by confusing signals from other reactors.  
But more importantly, perhaps, since one has in any real detector a limited resolution in 
energy and hence a limited resolution in $L/E$.  A typical value for this from KamLAND is 
about $\sigma_E / E = 6.2\%/\sqrt{E}$ in the positron energy.  This is limited in practice by 
the light emission from the scintillator and the amount of photocathode coverage of the 
detector.  While this is not an intrinsic physics limitation, improving upon this resolution 
by much more than a factor of two in a large instrument is probably not practical (with 
present technology).  Hence at a distance of order 100 km and typical neutrino energy of 3.5 
MeV, one would find the resolution to be about the same as the periodicity in L/E, and the 
desired modulation washed out.

At smaller distances, there are more events due to the 1/distance$^2$ dependence in the 
neutrino flux, but one intercepts fewer cycles in the $\Delta_{13}$ modulation.  This may not 
be a problem, witness the experiments (Double Chooz, Daya Bay, etc.) planned for the first 
dip at 1-2 km from reactors.  However, while one may use the expected (indeed need to use) 
shape of the spectral distortion for confirmation of any positive detection of finite 
$\theta_{13}$ in the close-in case, one has no substantial resolution in period of the 
oscillations and one is beholden to precise reactor neutrino spectral prediction and energy 
and background dependent detector systematics.  We are certainly not suggesting that such 
experiments will not work, nor even that they are not the best approach, but pointing out the 
limitations wherein a supplementary measurement as we describe herein may clinch the case in 
a largely independent measuring scheme, one which is self-normalizing (and yields other 
information, periods, into the bargain).

%%%%%%%%%%%% FIGURE 5 %%%%%%%%%%%%%%%
\begin{figure}[htbp]
%\vglue -0.50cm
\begin{center}
%\hglue  -0.5cm
\includegraphics[width=0.5\textwidth]{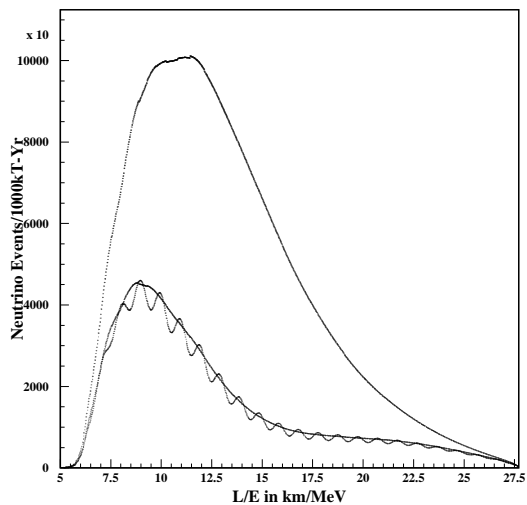}
\end{center}
%\vglue -1.8cm
\caption{Rate plotted versus L/E, in units of km/MeV, for the three cases as above.}
\label{loe}
\end{figure}
%%%%%%%%%%%% FIGURE 5 %%%%%%%%%%%%%%%

The new approach which we are proposing in this report is to utilize Fourier transform 
methods to extract the signal due to non-zero $\theta_{13}$, and along with this make several 
other measurements more precise.  We have calculated the expected spectrum of events at 
various distances and with varying values of $\sin^2(2\theta_{13})$. (Hereafter we will stick 
to the latter form for numerical values, rather than $\tan(\theta_{13})$ or 
$\sin^2(\theta_{13})$ or $\theta_{13}$, which are all used by various authors).

\begin{figure}[htbp]
%\vglue -0.50cm
\begin{center}
%\hglue  -0.5cm
\includegraphics[width=0.5\textwidth]{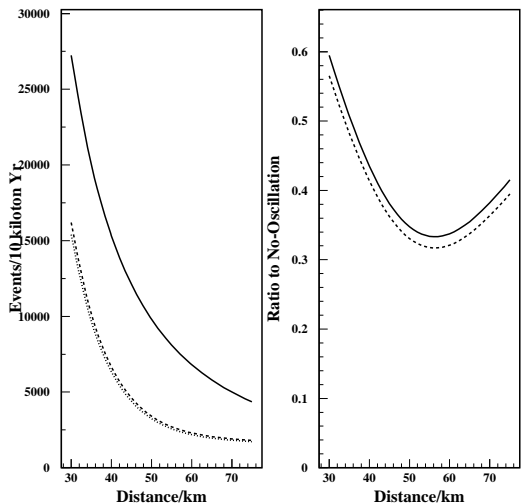}
\end{center}
%\vglue -1.8cm
\caption{Distance dependence for the cases of no-oscillations, oscillations without and with 
$\sin^2 (2\theta_{13}) = 0.1$.  The distance is from 30 to 60 km from the San Onofre complex and 
employing a 10 kiloton detector. The left hand panel shows the rates, while the right hand panel 
shows the suppression due to oscillations. Note maximal suppression at 50 km range. Also note 
difference in total rates due to $\theta_{13}$.}
\label{rate_vs_dist}
\end{figure}

We have simulated an event sample for a 10 kiloton fiducial volume detector operating for one 
year, and done most of our studies for a range of 50 km from the reactor. For definiteness we 
take the complex to have both reactors operating with fresh fuel, and the detector off shore 
from the San Onofre complex in southern California.  It would not make much difference what 
reactor complex we chose, as long as it is reasonably isolated and the thermal power is 
large, 6 GWt or more.  The energy spectrum was already shown in Figure \ref{spectrum}, and 
the distribution in L/E space in Figure \ref{loe}, which amounts to a total of 3472 
reactor-caused inverse $\beta$ events per year with $\theta_{13} = 0.0$.  The event rate 
versus distance is shown in Figure \ref{rate_vs_dist}, where one sees that the rate with 
oscillations falls at first faster than $1/r^2$ due to the main $\theta_{12}$ oscillation 
depleting the peak of the reactor spectrum.

\begin{figure}[htbp] 
%\vglue -0.50cm 
\begin{center} 
%\hglue -0.5cm 
\includegraphics[width=0.5\textwidth]{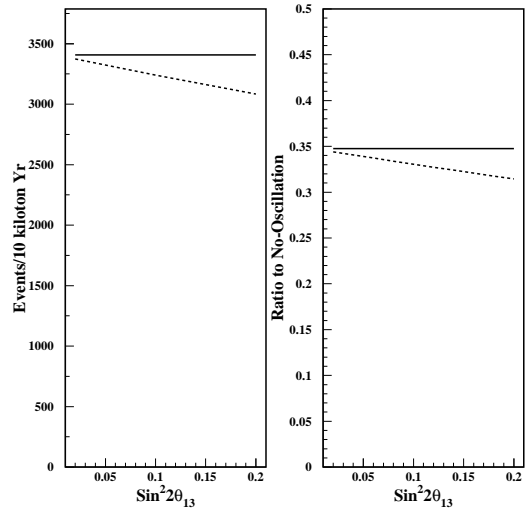} 
\end{center} 
%\vglue -1.8cm 
\caption{Dependence of the event rate on $\sin^2 (2\theta_{13})$. The distance is 50 km from the San 
Onofre complex, employing a 10 kiloton detector for 1 year. The left hand panel shows the rates, 
with and without $\theta_{13}$, while the right hand panel shows the suppression factor due to 
oscillations. Note 10\% effect on total rates due to $\theta_{13}$.} 
\label{rate_vs_s22t13} 
\end{figure}

In Figure \ref{rate_vs_s22t13} we show a similar plot scanning over values of 
$\sin^2(2\theta_{13})$ from 0 to the maximum presently allowed value of 0.2. Again this is 
for a range of 50km and 10 kiloton years exposure. An important observation here is that the 
total rate does depend upon $\theta_{13}$, and decreases with increasing mixing angle, with a 
total possible effect of about 10\%.  Hence, one cannot make a precision measurement of 
$\theta_{12}$ without knowing $\theta_{13}$.  This comes about through the dependence on 
$\cos^4(\theta_{13})$ term in the survival probability, Equation \ref{eqn:Pee}.

%%%%%%%%%%%% FIGURE 6 %%%%%%%%%%%%%%%
%%%%%%%%%%%% FIGURE 6 %%%%%%%%%%%%%%%

\begin{figure}[htbp]
%\vglue -0.50cm
\begin{center}
%\hglue  -0.5cm
\includegraphics[width=0.5\textwidth]{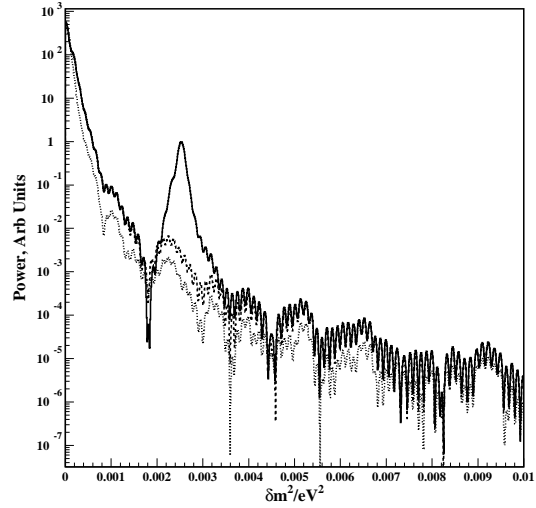}
\end{center}
%\vglue -1.8cm
\caption{Power spectrum for the cases of no-oscillations, oscillations with $\sin^2 (2\theta_{13}) 
= 0.1$, and without. The modulation is in units of $eV^2$ and power in arbitrary units on the 
logarithmic vertical axis, normalized to zero $\delta m^2$. The distance is 50 km from the San 
Onofre complex. Note distinctive peak at $(\delta m^2)_{13}$, well above background. The low 
frequency peak contains both ${\delta m^2}_{solar}$ modulation and the effective envelope of the 
reactor spectrum.}
\label{logpwr}
\end{figure}

Next we show the Fourier transform of the expected data, employing a 1000 bin discretization 
in L/E space, as shown in Figure \ref{logpwr}.  This spectrum (in $\delta m^2$ space, which 
one may think of as frequency as compared to the equivalence of L/E space with time) is 
dominated, of course, by the major low frequency peak.  One may think of the energy spectrum 
as giving us the ``envelope function''.  Nice envelope functions go smoothly to zero on the 
extremes, and suppress sidelobes in the transform (in contrast to a flat L/E distribution 
which would have square-wave type sidelobes).  Unfortunately (as is obvious) the major 
$\theta_{12}$ oscillations having only roughly one (or less) cycle in the energy "bandwith" 
yield only a low frequency lump in the transform.  Hence, Fourier transforms are not useful 
for analyzing the $\theta_{12}$ oscillations: it will simply have to be fitted, as is 
normally done. However, as we saw earlier in Figure \ref{red}, the lines of constant L/E 
become closer together in energy with distance, and one intercepts more cycles in the 
$\theta_{13}$ oscillation.  Hence the resolution of the peak improves with distance, though 
in competition with falling statistics.

%%%%%%%%%%%% FIGURE 6' %%%%%%%%%%%%%%% 
%%%%%%%%%%%% FIGURE 6' %%%%%%%%%%%%%%%

We do not put in background and and do not make a cut for geoneutrinos in this study, though 
this must be accounted for in later work.  We note though that the results for $\theta_{13}$ 
will be quite insensitive to continuous spectrum backgrounds which will only contribute to 
the low frequency peak.  This is part of the beauty of the Fourier approach, since it picks 
out only the desired oscillations, and is insensitive to the overall shape of the neutrino 
spectrum.

%%%%%%%%%%%% FIGURE 7 %%%%%%%%%%%%%%%
%%%%%%%%%%%% FIGURE 7 %%%%%%%%%%%%%%%

\begin{figure}[htbp]
%\vglue -0.50cm
\begin{center}
%\hglue  -0.5cm
\includegraphics[width=0.5\textwidth]{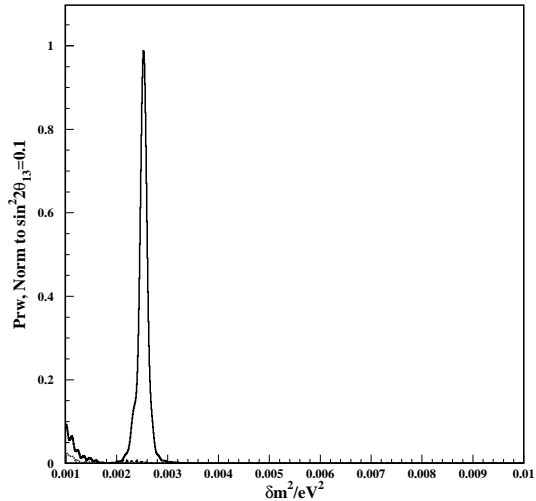}
\end{center}
%\vglue -1.8cm
\caption{Power spectrum around $\theta_{13}$ peak, on a linear scale. The cases with 
no-oscillation and no signal are barely discernable at the bottom, due to linear scale 
in power here. Note lack of sidebands or aliases.}
\label{pwrpeak}
\end{figure}

We show in Figure \ref{pwrpeak} the power spectrum in the neighborhood of the $\theta_{13}$ 
peak, and one sees what a nicely separated and clean signal is evident (for $\sin^2 
(2\theta_{13})$ = 0.1).  The null case is also plotted, which is to say the case of 
$\theta_{13} = 0.0$, with the same statistics (10 kiloton years at 50 km.), and is nearly 
invisible.  All spectra shown have been smeared in positron energy with a resolution function 
of $\delta E = 0.032/\sqrt{E_{positron}}$.

%%%%%%%%%%%% FIGURE 8 %%%%%%%%%%%%%%%
%%%%%%%%%%%% FIGURE 8 %%%%%%%%%%%%%%%

\begin{figure}[htbp]
%\vglue -0.50cm
\begin{center}
%\hglue  -0.5cm
\includegraphics[width=0.5\textwidth]{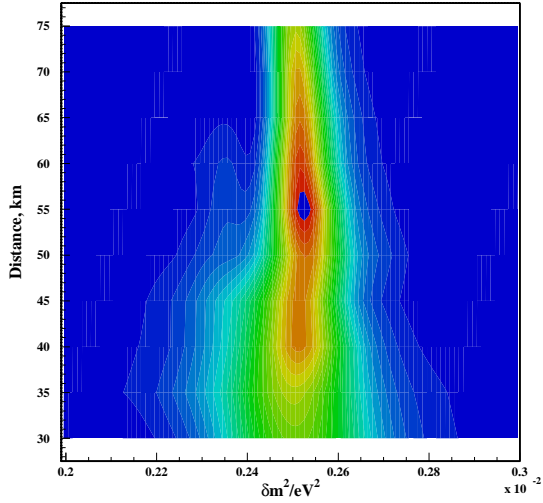}
\end{center}
%\vglue -1.8cm
\caption{Power spectrum near the $\theta_{13}$ peak versus distance from the power station, 
with $\sin^2 (2\theta_{13}) = 0.1$. Note the peaking near 50 km, due to increasing peak 
spread at lesser distances and smearing of oscillations at larger distances.}
\label{t13_peak_vs_dist}
\end{figure}

We show the shape of the $\theta_{13}$ peak versus distance in Figure \ref{t13_peak_vs_dist}, 
and one sees indeed that the peak spreads at distances nearer than 50 km, while dropping off 
at greater distances due to falling statistics.  This appears to be as we expected since in 
this region the L/E distribution is reasonably flat, and we encompass about a dozen cycles.

%%%%%%%%%%%% FIGURE 10 %%%%%%%%%%%%%%%
%%%%%%%%%%%% FIGURE 10 %%%%%%%%%%%%%%%

\begin{figure}[htbp]
%\vglue -0.50cm
\begin{center}
%\hglue  -0.5cm
\includegraphics[width=0.5\textwidth]{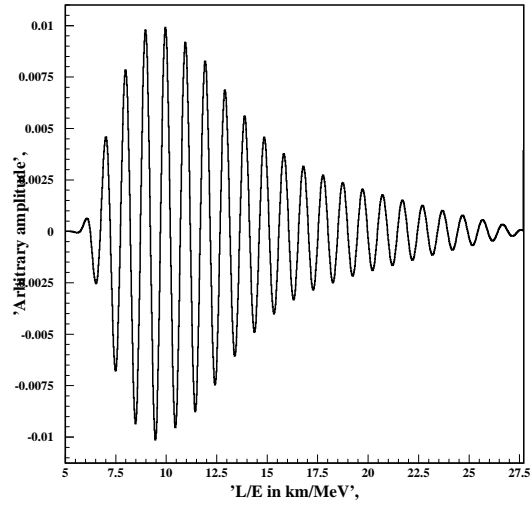}
\end{center}
%\vglue -1.8cm
\caption{Matched filter function at 50km. }
\label{filter_function}
\end{figure}

\begin{figure}[htbp]
%\vglue -0.50cm
\begin{center}
%\hglue  -0.5cm
\includegraphics[width=0.5\textwidth]{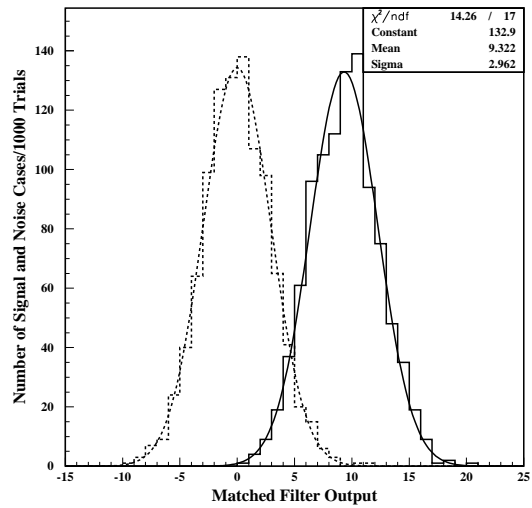}
\end{center}
%\vglue -1.8cm
\caption{Matched filter output for the cases of no signal and 
${\sin}^2(2\theta_{13}) = 0.1$, for 1000 trials at 50 km. }
\label{filter_spread}
\end{figure}

A more powerful approach than using the Fourier power spectrum, since we know the expected 
phase and can determine the peak `frequency' with high accuracy, is to employ a ``matched 
filter'' function to convolve with the data in the time domain (L/E).  The filter function is 
shown for 50 km in Figure \ref{filter_function}, and the results are indicated in Figures 
\ref{filter_spread}, at 50 km distance.

%%%%%%%%%%%% FIGURE 14 %%%%%%%%%%%%%%%
%%%%%%%%%%%% FIGURE 14 %%%%%%%%%%%%%%%

%%%%%%%%%%%% FIGURE 13 %%%%%%%%%%%%%%%
%%%%%%%%%%%% FIGURE 13 %%%%%%%%%%%%%%%

\begin{figure}[htbp]
\begin{center}
\includegraphics[width=0.5\textwidth]{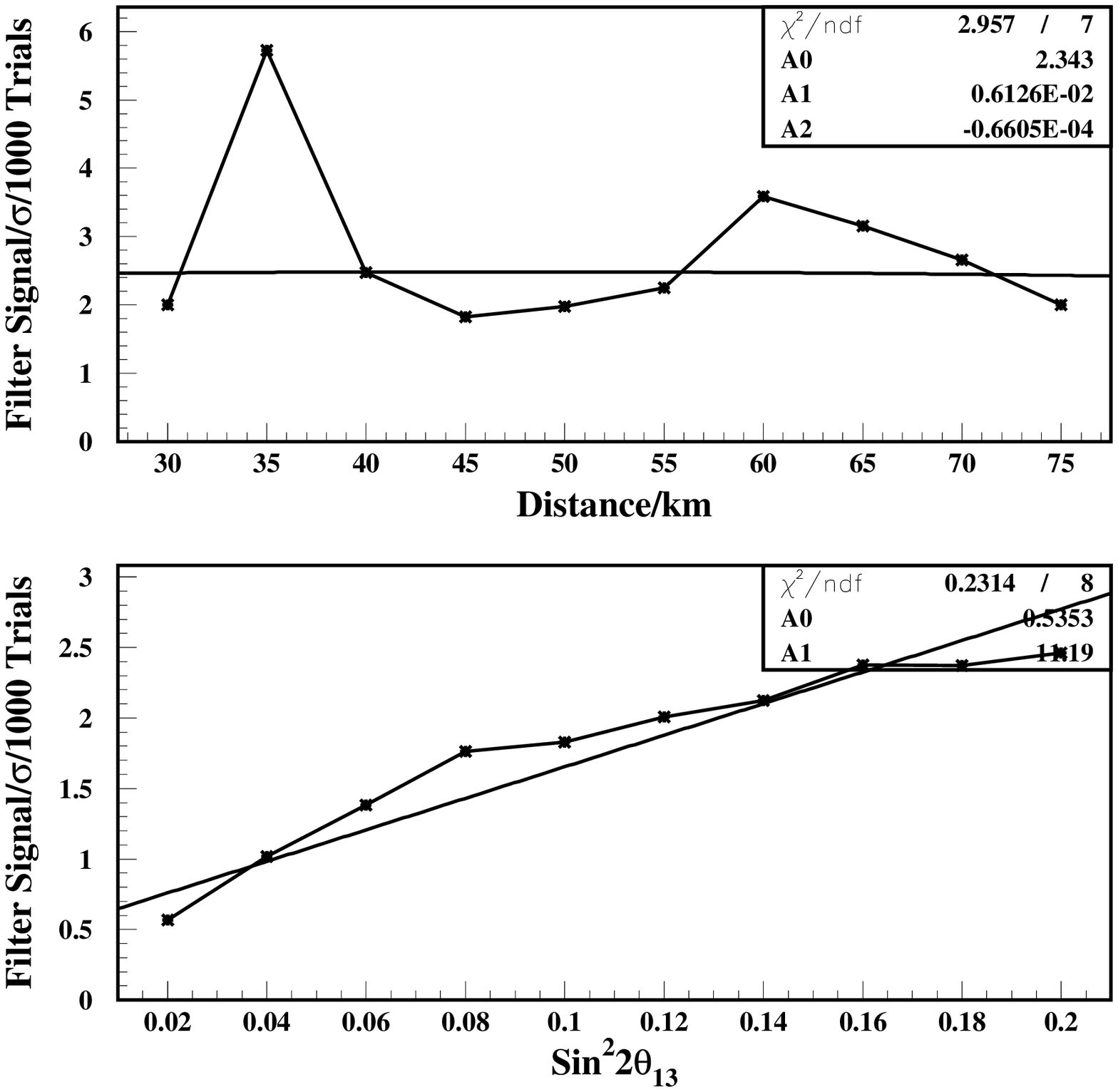}
\end{center}
\caption{Matched filter $signal/\sigma_{signal}$ versus distance in
 km and $\sin^2(2\theta_{13})$.} 
\label{filter_signifs}
\end{figure}

Finally we show the results of matched filter scanning of 1000 simulated data sets at each of 
10 different distances and 10 different values of $\sin^2(2\theta_{13})$ in Figure 
\ref{filter_signifs}, for an exposure of 10 kiloton years. One would think that it might be 
better to conduct the measurements at distances of less than 50 km, but that ignores the peak 
spread at smaller distances, and greater sensitivity to background.  The significance varies 
with the square root of exposure. We estimate that at 50 km, one could reach down to 
$\sin^2(2\theta_{13}) = 0.03$ after a 100 kiloton-yr exposure at 3 $\sigma$.

%%%%%%%%%%%% FIGURE 16 %%%%%%%%%%%%%%%
%%%%%%%%%%%% FIGURE 16 %%%%%%%%%%%%%%%

%%%%%%%%%%%%%%%%%%%%%%%%%%%%%%%%%%%%%%%%%%%%%%%%%%%%%%%%%%%%%%%%%%%%%%%%%%%
\subsection{Determining the mass hierarchy}
\label{hierarchy}
%%%%%%%%%%%%%%%%%%%%%%%%%%%%%%%%%%%%%%%%%%%%%%%%%%%%%%%%%%%%%%%%%%%%%%%%%%%

Using this technique it is possible to determine the neutrino mass hierarchy by resolving the 
asymmetry of the transform due to the small shoulder displaced by $\delta {m^2}_{12}$ from 
the main peak. The shoulder with a power reduced by about a factor of 6 is at smaller $\delta 
m^2$ for normal hierarchy and at larger $\delta m^2$ for inverted hierarchy. In order to 
assess the quantitative ability of an experiment to discriminate between normal and inverted 
hierarchy, we have written a simulation program which generates and analyzes data sets from 
an idealized $8.5 \times 10^{32}$ free proton detector and 6 GW$_{th}$ reactor complex. We 
have varied the range, mixing angle ($\sin^2(2\theta_{13})$), and exposure time, making 1000 
simulated experiments at each set of parameters.

We have not at this stage included detector specific background sources such as those due to 
cosmic ray muons traversing the detector, radio impurities, geophysical neutrinos, or 
neutrinos from other (more distant) reactors. The cosmic ray induced background depends upon 
depth of water or rock overburden, so must be assessed for the individual proposed location. 
We know, however that this is of no concern at depths greater than 3 kmwe, though lesser 
depths may be acceptable. Other reactors will make a small contribution, if sites are chosen 
on the basis of not having significant additional flux (though to a certain extent these can 
be included in the analysis). In general we do not expect background to compromise the 
proposed method, since the added neutrinos start at random distances relative to the 
detector, so make no coherent contribution to the Fourier transform on L/E at the "frequency" 
of interest. One may think of such background, if uniformly distributed in L/E as simply 
contributing to the zeroth term in the transform, the total rate. Of course, the more random 
events in a finite sample, the more background across the $\delta m^2$ spectrum. In any 
event, at this stage we neglect background, reserving the study for more specific 
applications.
  
We have studied several algorithms for determining the mass hierarchy, noting that the 
periodicity ($\delta m^2$), if evident, is measured to 0.1\% precision. In practice this is 
limited by systematic uncertainties in terms of interpretation as a particular mass 
difference, probably the energy scale uncertainty (of order 1\%). However, in the data set 
the peak is known to whatever we fit it to, and we can analyze the data employing that 
knowledge. Hence, knowing the primary peak ($\delta {m^2}_{13}$), we need to determine if the 
secondary peak is at greater or lesser values of $\delta m^2$. While we do not know $\delta 
{m^2}_{12}$ exceedingly accurately, we know $\sigma_{12}/\delta {m^2}_{13}$ very well (to 
about $3\times 10^{-3}$). This is to be compared to the spread of about 3\% between the two 
peaks. Hence we can construct a measure examining how well the data fit each hierarchy 
hypothesis. For presentation here, we use a "matched filter" approach, which one can think of 
as the Fourier transform of the correlation function, producing a numerical value for each 
hypothesis.

\begin{figure}[htbp]
\begin{center}
\includegraphics[width=0.5\textwidth]{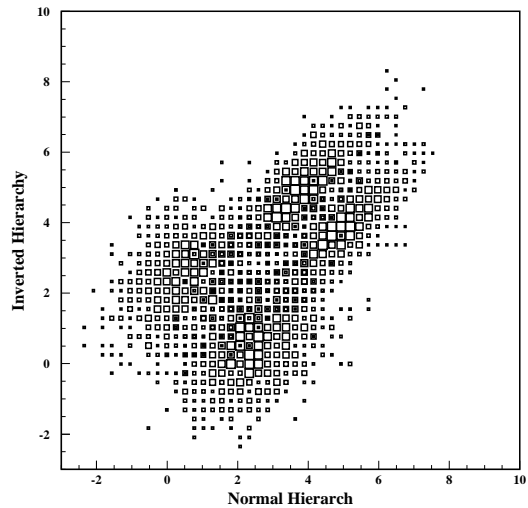}
\end{center}
\caption{Distance dependent scatter plots for hierarchy test. The plots on the lower right are sets 
of 1000 experiments at 30 and 50 km with normal hierarchy. Those on the upper left are with 
inverted hierarchy. }
\label{fig:disdis}
\end{figure}

\begin{figure}[htbp]
\begin{center}
\includegraphics[width=0.5\textwidth]{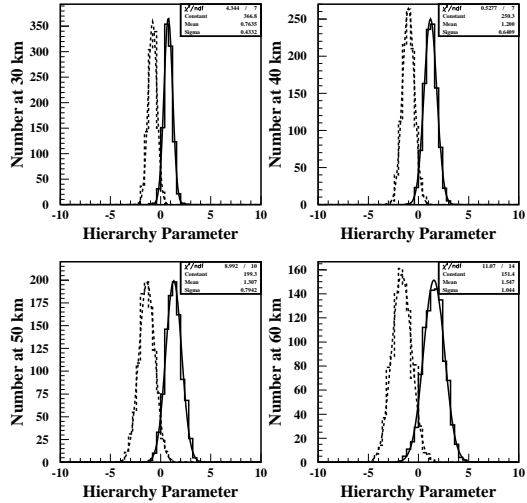}
\end{center}
\caption{Hierarchy parameter distributions for 30, 40, 50, and 60 km. Solid histograms are with 
normal hierarchy, dashed with inverted. Distributions fit well to a Gaussian. }
\label{fig:dispar}
\end{figure}

In Figure \ref{fig:disdis} we show in a scatter plot the distribution of "experimental" results at 
distances of 30, 40, 50, and 60 km with normal and inverted hierarchy. Each experiment yields two 
numbers, the output of the matched filter, which we plot on the x and y axes. One sees that there 
is very nice separation along the diagonal. Hence we construct a new variable by projecting the 
distributions onto a 45 degree line. This is illustrated in Figure \ref{fig:dispar} in four panels. 
The data fits well to a Gaussian distribution. Separation is quite good ($>$95\%) over the entire 
range examined, from 30-75 km, but falls off below 40 km and above 65 km.
  
\begin{figure}[htbp]
\begin{center}
\includegraphics[width=0.5\textwidth]{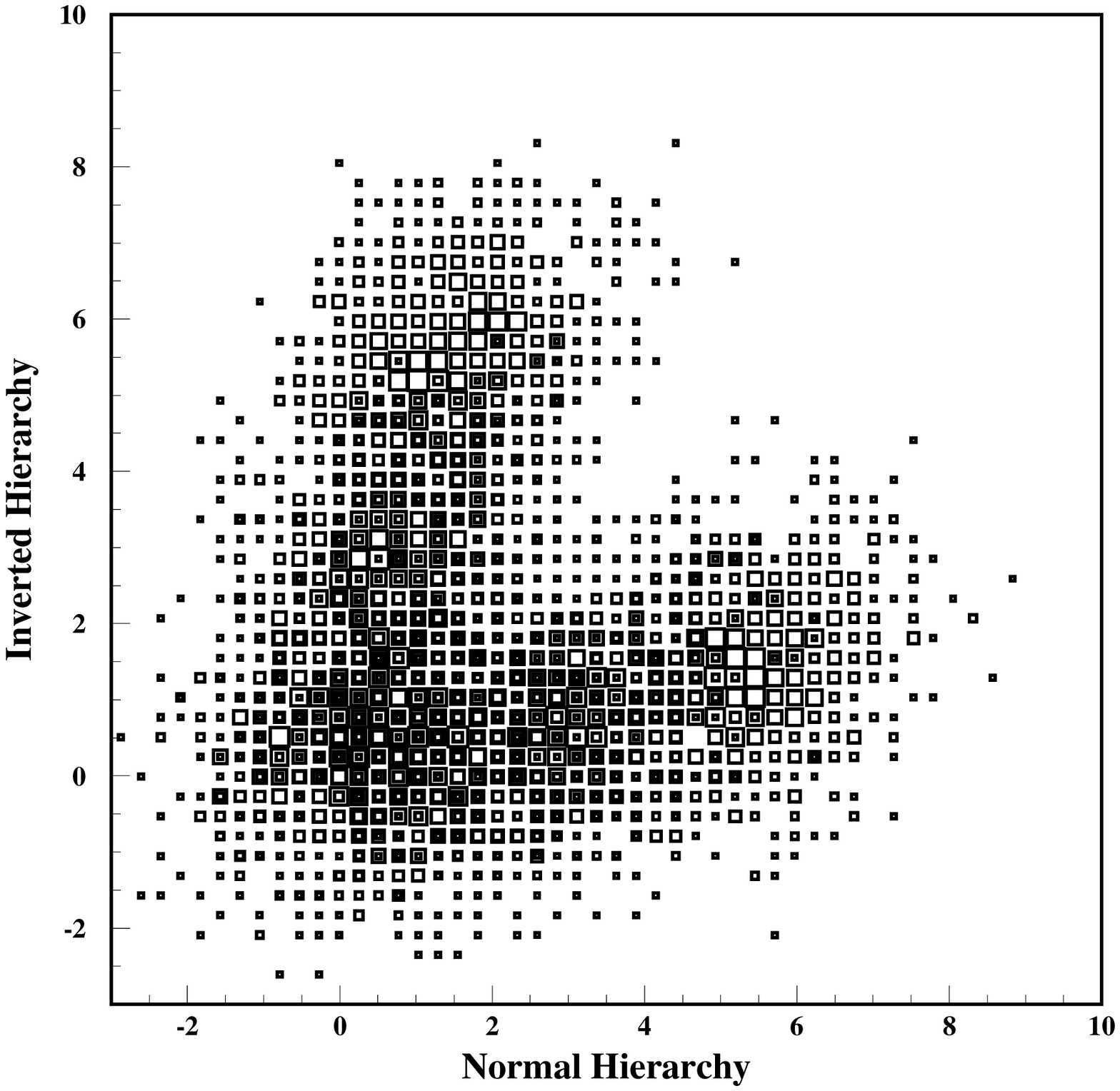}
\end{center}
\caption{$\sin^2(2\theta_{13})$ dependent scatter plots for hierarchy test using matched filter 
output. Horizontal plots are sets of 1000 experiments at $\sin^2(2\theta_{13})$ = 0.04, 0.12, and 
0.20 with normal hierarchy. Vertical plots are with inverted hierarchy. Note the greater separation 
with larger $\sin^2(2\theta_{13})$ }
\label{fig:s22dis}
\end{figure}

\begin{figure}[htbp]
\begin{center}
\includegraphics[width=0.5\textwidth]{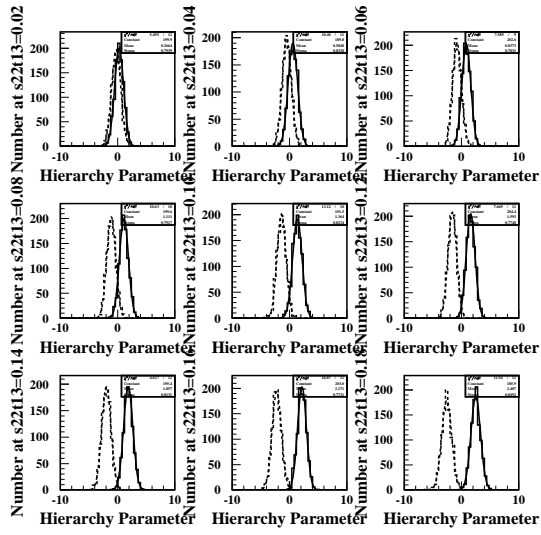}
\end{center}
\caption{Hierarchy parameter distributions for 1000 experiments each with $\sin^2(2\theta_{13})$ 
values of 0.02, 0.04, 0.06, 0.08, 0.10, 0.12, 0.14, 0.16, and 0.18. Solid histograms are with 
normal hierarchy, dashed with inverted.}
\label{fig:s22par}
\end{figure}

Next we examine the sensitivity of the hierarchy determination to $\sin^2(2\theta_{13})$. In 
Figure \ref{fig:s22dis} we present scatter plots of hierarchy tests for 1000 experiments at 
each of $\sin^2(2\theta_{13})$ = 0.04, 0.12 and 0.20, all at 50 km range. One sees that the 
distributions are well separated at $\sin^2(2\theta_{13})$ values more than about 0.04 in one 
year). The values of the hierarchy parameter are plotted in the same projection as above for 
the distance study, in Figure \ref{fig:s22par}. It thus appears as though such an experiment 
can probe the hierarchy down to $\sin^2(2\theta_{13})$ values of 0.02 with an exposure of 100 
kT-y (with the caveats about site specific background).

%%%%%%%%%%%%%%%%%%%%%%%%%%%%%%%%%%%%%%%%%%%%%%%%%%%%%%%%%%%%%%%%%%%%%%%%%%%
\subsection{Measuring $\Theta_{12}$}
\label{t12}
%%%%%%%%%%%%%%%%%%%%%%%%%%%%%%%%%%%%%%%%%%%%%%%%%%%%%%%%%%%%%%%%%%%%%%%%%%%

We would like to touch briefly on a precision measurement of $\theta_{12}$ with this detector 
(Briefly, not because it is not important, but because it has been widely discussed 
elsewhere, and surely represents a major and straight forward goal).  The idea of making a 
precision measurement of $\theta_{12}$ by placing a detector at the baseline corresponding to 
an oscillation maximum for the peak reactor neutrino energy has been the subject of some 
discussion recently\cite{3nu1,3nu2}.  The optimum distance for this measurement has been 
claimed to be 50-70 km, and it has been argued that for an exposure of about 60 GWt-kT-Yr a 
measurement of $\sin^2(\theta_{12})$ to about 2\% (at 1 $\sigma$) can be made.  For a 
detector of 10 kT fiducial mass, such as we are considering, obviously one can do better.

For measuring $\theta_{12}$ one cannot go far enough from the reactor to get many cycles in 
the solar oscillation period within the reactor energy bandwidth. Working at about the range 
already identified (50-60 km) one will get major suppression across the middle of the reactor 
spectrum, as illustrated in Figures \ref{red} and \ref{spectrum}. The logic for measuring at 
this distance depends upon utilization of predicted neutrino spectra and cross sections, as 
well as detector systematics, such as precise knowledge of the fiducial volume (which has 
been a limiting factor in KamLAND).  We think that there may be another way to go about this 
quest, however, in a experiment free of much of the systematics, if one moves even as far 
away as to 100 km.  Going twice as far away costs a factor of four in event rate, but one 
recovers almost half of this by being near the major oscillation's first return for 3.5 MeV 
neutrinos.  Here one fits the rapidly decreasing event rate with energy to achieve an 
independent value of the mixing angle, the measurement being largely self-normalizing and 
free of systematics such as fiducial volume. The major remaining systematic uncertainty in 
this instance will be in the relative spectral shape of the reactor flux as it falls from 3.5 
to 6.0 MeV.  At the one percent level this should be a manageable problem, particularly if 
one is taking into account the fuel state of the reactors.

We remain uncommitted about the optimal distance for making a precise $\theta_{12}$ 
measurement, but we at least claim that 50-60 km will do well.  We leave it to a later report 
which must include practical matters such as available depths versus distance, neighboring 
reactors, and a real background assessment, to make firm conclusions about exposures, 
distance and achievable resolution.  Note that the measurement of $\theta_{12}$ does not 
depend upon a non-zero value of $\theta_{13}$.

%%%%%%%%%%%%%%%%%%%%%%%%%%%%%%%%%%%%%%%%%%%%%%%%%%%%%%%%%%%%%%%%%%%%%%%%%%%
\subsection{Implications and Further Studies}
\label{implications}
%%%%%%%%%%%%%%%%%%%%%%%%%%%%%%%%%%%%%%%%%%%%%%%%%%%%%%%%%%%%%%%%%%%%%%%%%%%

Further work is needed to study the effects of background due to geoneutrinos and cosmic ray 
induced events.  The latter are strongly site and detector depth dependent, and so one needs to 
model specific depth profiles off-shore.  As stated earlier, we expect little influence of the extra 
events on the $\theta_{13}$ measurements, since these depend upon relatively high frequency ($\delta 
{m^2}_{13}$) variations over the spectrum.  For the $\theta_{12}$ measurement, they will need to be 
incorporated in the analysis. Note that the $\theta_{12}$ measurement we propose will be relaively 
free of systematics, even so.  Basically the fit for the depletion of the non-oscillated flux is 
compared to itself, and so is dominated by statistics, not by normalization to a near detector.  
This measurement is sensitive to the shape of the reactor spectrum, which is better known than the 
absolute rates (which include errors in cross-section, detector fiducial volume, energy independent 
efficiency and reactor power).

Another area for exploration involves splitting the observation into two or more distances.  We have 
in mind the possibility to employ some time at, say 50 km and some time at perhaps 100 km.  The 
virtue of this would be the ability to co-analyze the data sets, incorporating more cycles in 
$\theta_{13}$, and perhaps two cycles in $\theta_{12}$.  The data can be added in L/E space prior to 
Fourier transform since we know the L and E well enough (L to $10^{-4}$, E to perhaps $10^{-3}$ in 
possible systematic shift with same detector, different distance).  Hence the Fourier power will 
add, and contribute as the square of events in the overlap region.

We also need to study the value of a close-in detector, as in the case of San Onofre, which already 
has a nearby monitor in place\cite{songs}.  Of course this detector cannot make a precision 
prediction of rates for the distant and much larger Hanohano type of instrument which will have 
very different efficiencies and systematics. But, the near detector should provide long term stable 
recording of the reactor neutrino output (as opposed to the present dependence upon thermal output 
measurments), as well as monitoring of nuclear fuel aging.

%%%%%%%%%%%%%%%%%%%%%%%%%%%%%%%%%%%%%%%%%%%%%%%%%%%%%%%%%%%%%%%%%%%%%%%%%%%
\subsection{Concluding Remarks about the Physics Prospects for Hanohano}
\label{conclusion}
%%%%%%%%%%%%%%%%%%%%%%%%%%%%%%%%%%%%%%%%%%%%%%%%%%%%%%%%%%%%%%%%%%%%%%%%%%%

We have shown that a measurement of the spectrum of anti-neutrino events at distances of tens 
of kilometers from a strong reactor source can yield a significant measurement of the two 
important neutrino mixing angles, $\theta_{13}$ and $\theta_{12}$.  The latter can be 
accomplished with a 10 kiloton scale liquid scintillation detector with an exposure of order 
of a year, if the $\sin^2 (2\theta_{13})$ is not smaller than about 0.05. Longer exposures 
and larger detector can of course reach smaller values.

One of the attractions of such a measurement is that it does not depend upon normalization to 
a close-in detector, nor on difficult measures to achieve very small systematic errors.  We 
see such a measurement not as a competition to the proposed experiments such as Double Chooz, 
Daya Bay, and Kaska but as a compliment to confirm their results in the case of a signal in 
the range above about 0.05. A next generation instrument of 100 kilotons and 5 years exposure 
should be able to reach $\sin^2 (2\theta_{13}) \simeq 0.01$, for example.

Moreover, the measurements described herein, employing Fourier transform and signal 
processing techniques, have the ability to extricate precise determinations of mass 
differences, as we have described.  More exciting is the prospect for determining the 
neutrino mass hierarchy with a method which does not depend upon matter effects and which is 
not very sensitive to systematic errors.

\subsection{Other Particle Physics and Astrophysics}

Recall that Hanohano will be 20 times the mass of KamLAND, ten times SNO+, 50 times Borexino.  
Because of the deep deployments for geonu measurements Hanohano will also have lower 
background than other instruments.  And the rare astrophysical events with be observed 
completely without interference with the geonu studies.

Perhaps the most exciting prospects are for supernova neutrinos. SuperK and the proposed next 
generation of Water Cherenkov detectors will have greater fiducial volume for detecting 
supernova neutrinos in the energy range of typically 10-50 MeV for a galactic supernova.  
But the liquid scintillation detector has the unique ability to measure the electron 
anti-neutrino content of the flux. Proton recoils may also be detected, a unique measurement.
Charged current and neutral current events on C12 are also important.

Another subject of much interest is the detection of the sum of neutrinos from all past 
supernovae at all distances.  At great enough distance, the energies appear degraded to us 
due to the universe expansion and so the spectrum is shifted downwards. This flux of relic SN 
anti-neutrinos has not yet been detected, but Hanohano will be in a good position to do so, 
due to the lack of background (which limits detection in a Water Cherenkov instrument sensing 
neutrinos).

Another topic of interest, which will proceed without interference during other observations with 
Hanohano, is the search for nucleon decay, one of the greatest challenges of particle physics, and 
directly bearing upon theories of grand unification.  While the larger Water Cherenkov detectors 
once again will hold the field in sheer size, the low energy threshold of the scintillation 
detector makes possible some unique probes. In particular, because of the large mass of the kaon, 
proton decay into kaon containing modes is not efficient in the large Water Cherenkov detectors, 
but is so in a scintillation instrument. At present the $\tau/b~>~2.3 \times 10^{33}~ y$ [Super-K: 
PR D 72, 052007 (2005)]. Hanohano can reach a $\tau/b~>~10^{34}~y$ with a 10 yr run. [Lena: PR D 
72, 075014 (2005)] Also Hanohano can make record searches for neutron disappearance.  At present 
the limit is $\tau(n \rightarrow invisible modes)~ > ~5.8 \times 10^{29}~y$ at 90\% CL and $\tau(nn 
\rightarrow invisible modes)~ >~ 1.4 \times 10^{30}~y$ at 90\% CL [838 and 1119 metric ton-years of 
KamLAND: PRL 96 (2006) 101802] Hanohano may achieve $\tau(n \rightarrow invisible modes)~ >~ 5 
\times 10^{31}~ y$ at 90\% CL 10 yrs and $\tau(nn \rightarrow invisible modes)~ >~ 5 \times 
10^{31}~ y$ at 90\% CL.

As usual with instruments breaking into a new regime in size and sensitivity by an order of magnitude or 
more, Hanohano will have the opportunity to detect unusual phenomena.  There is a long list of limits on 
exotica such as magnetic monopoles, quark nuggets, micro-black holes, etc., for which Hanohano will have 
new serendipitous potential.

\section{Other Applications of Future Large Low Energy Neutrino Detectors}

In the future we can anticipate many uses of neutrinos both for fundamental science in particle 
physics and astrophysics, and in applications as probes due to their unique penetrating ability.  
For some time now people have written papers suggesting some far-out possibilities, such as active 
earth tomography with accelerator produced neutrino beams and perhaps natural neutrinos, using 
neutrino beams to search for oil, measure heterogeneities, measuring earth core properties in ways 
unrivaled, and even as carrier beam for an ultimate galactic time standard.

In the shorter term we can begin to think seriously about using neutrinos to monitor nuclear 
reactors, both for checking on use of the reactor and reactor performance.  This can only 
reasonably be carried out from close-in (10-100 m) and with cooperative facilities. For 
locations which may not be cooperative, one can stand away distances of hundreds or even 
thousands of kilometers.  However, the price for larger range is greater detector volume, of 
course, since the flux falls with distance squared. Moreover one will start to have competing 
signals from other reactors.  Since there are about 500 reactors in the world, one can 
imagine a network of roughly that number of detectors which can monitor all the world's 
reactors, and can subtract the known contributions from cooperative sites, revealing 
clandestine operations.  While there are other means to search for rogue reactors (e.g. 
thermal signatures), one cannot shield the neutrinos.  And, the synergistic application of 
multiple monitoring techniques may yield more powerful constraints\cite{nacw}.

Another application in this line, which comes for free with remote (close-in detectors would 
not have the sensitivity) reactor monitoring is the detection of clandestine nuclear weapons 
testing.  Again, there are many mechanisms in place for detection of such activity, there 
have been cases of both false positives and false negatives.  The neutrino signature cannot 
be faked or masked, and is a definitive measure of the weapons fission yield.  Studies have 
shown that a large array for reactor monitoring as above, could detect weapons down to the 
one kiloton level anywhere in the world.

Science applications of future huge low energy neutrino detectors are also very exciting.  
For example, a one gigaton instrument (or collection totaling that effective mass) could 
detect supernovae from throughout our galactic supercluster, recording perhaps one per week.  
Such would have many associated studies ranging from stellar evolution to general relativity 
and particle physics.  The measurement of the sum of electron anti-neutrinos from all 
previous supernovae throughout the universe would also yield much interesting information 
upon stellar formation rates and cosmology.  On a more local level, the increased thermal 
neutrino output of a star in the last few days of burning prior to implosion may be 
registered with large instruments from throughout our galaxy, giving earth a supernova early 
warning system\cite{doanow}.

\section{Conclusion}

In summary, in one year of observations with Hanohano we can achieve the following measurements:

{\bf Neutrino Geophysics, deep mid-ocean}

	Mantle flux U/Th geo-neutrinos to $\simeq$25\%

	Measure Th/U ratio to $\simeq$20\%

	Rule out geo-reactor of P$>$0.3 TW

	Annual changes in location can begin study of lateral heterogeneity of U/Th

{\bf Neutrino Particle Physics, 50 km from reactor}

	Measure $sin^2(\theta_{12})$ to few \% with 1/2 cycle observation

	Measure $sin^2(2\theta_{13})$ down to $\simeq$0.05 w/ multi-cycle observations

	Measure $\delta {m^2}_{31}$ at percent level w/ multi-cycle

	No near detector; insensitive to background, systematics; complimentary to Double 
          Cooz, Daya Bay, Minos, Nova 

	Determine mass hierarchy, depending upon $\theta_{13}$

The study of a deep ocean electron anti-neutrino detector has evolved into a plan for an 
experiment which can attack major scientific questions in both geology and particle physics.  
It represents a start into a new scientific area, and aims toward future practical neutrino 
applications.  We are very excited about the prospects for this program, and look forward to 
this new adventure.

\section{Acknowledgements}
  
Many have contributed to the material presented herein. Joe Van Ryzin of Makai Ocean 
Engineering has been the leader of the engineering studies.  Bob Svoboda of UC Davis and LLNL 
helped with calculations of reactor fluxes.  Gene Guillian now of Queens University, Canada 
and Jelena Maricic have done extensive calculations on reactor monitoring and natural 
neutrino fluxes. Thanks also to colleagues Gary Varner and Shige Matsuno, and students 
Mavourneen Wilcox and Peter Grach. Our geology colleagues, in particular Bill McDonough of U. 
Md., have done a great deal in getting this program under way.  We draw heavily on the work 
of our friends and colleagues in the KamLAND, SNO, Borexino and LENA groups for the 
development and understanding of the type of instrument uner study.  We also acknowledge 
support from CEROS\cite{ceros} for much of the laboratory studies and preliminary design 
work.  We want to thank the organizers of the Venice Neutrino telescopes Conference series, 
and Prof. Milla Baldo-Ceolin in particular for her charm, taste, and steady hand in guiding 
this unparalled set of meetings.

\end{document}